\documentclass[twocolumn]{aastex631}

\usepackage{graphicx}  % Including figure files
\usepackage{amsmath}   % Advanced maths commands
\usepackage{amssymb}   % Symbols
\usepackage{mathrsfs}  % For script fonts
\usepackage{xcolor}    % Color support
\usepackage{ulem}      % Underlining
\usepackage{cancel}    % Strikeout
\usepackage{bm}        % Bold math symbols
\usepackage{subcaption}  % For subfigures (includes caption)
\usepackage{hyperref}  % Hyperlinks and references

\newcommand{\emre}{\color{black}}

\begin{document}

\title{On the feasibility of deriving pseudo-redshifts of gamma-ray bursts from two phenomenological correlations}

\author[0000-0002-8442-9458]{Emre S. Yorgancioglu}
\affiliation{Key Laboratory of Particle Astrophysics, Institute of High Energy Physics, Chinese Academy of Sciences \\
19B Yuquan Road, Beijing 100049, People’s Republic of China}
%\email{emre@ihep.ac.cn} 
\email{emre@ihep.ac.cn}

\affiliation{University of Chinese Academy of Sciences, Chinese Academy of Sciences, Beijing 100049, People’s Republic of China \\
}

\author[0000-0002-1905-1727]{Yun-Fei Du}
\affiliation{Key Laboratory of Particle Astrophysics, Institute of High Energy Physics, Chinese Academy of Sciences \\
19B Yuquan Road, Beijing 100049, People’s Republic of China}

\affiliation{University of Chinese Academy of Sciences, Chinese Academy of Sciences, Beijing 100049, People’s Republic of China \\
}

%\collaboration{20}{(AAS Journals Data Editors)}

\author[0000-0001-7599-0174]{Shu-Xu Yi $\dagger$ }
\affiliation{Key Laboratory of Particle Astrophysics, Institute of High Energy Physics, Chinese Academy of Sciences \\
19B Yuquan Road, Beijing 100049, People’s Republic of China}
\email{ $\dagger$ sxyi@ihep.ac.cn}

\author[0000-0002-2516-5894]{Rahim Moradi}
\affiliation{Key Laboratory of Particle Astrophysics, Institute of High Energy Physics, Chinese Academy of Sciences \\
19B Yuquan Road, Beijing 100049, People’s Republic of China}
\affiliation{INAF -- Osservatorio Astronomico d'Abruzzo,Via M. Maggini snc, I-64100, Teramo, Italy}

\author[0000-0001-7584-6236]{Hua Feng}
\affiliation{Key Laboratory of Particle Astrophysics, Institute of High Energy Physics, Chinese Academy of Sciences \\
19B Yuquan Road, Beijing 100049, People’s Republic of China}

\author[0000-0001-5586-1017]{Shuang-Nan Zhang $\ddagger$}
\affiliation{Key Laboratory of Particle Astrophysics, Institute of High Energy Physics, Chinese Academy of Sciences \\
19B Yuquan Road, Beijing 100049, People’s Republic of China}
\affiliation{University of Chinese Academy of Sciences, Chinese Academy of Sciences, Beijing 100049, People’s Republic of China \\
}
\email{ $\ddagger$ zhangsn@ihep.ac.cn}

%\thanks{\dagger}



\begin{abstract}
\noindent Accurate knowledge of gamma-ray burst (GRB) redshifts is essential for studying their intrinsic properties and exploring their potential application in cosmology. Currently, only a small fraction of GRBs have independent redshift measurements,  primarily due to the need of rapid follow-up optical/IR spectroscopic observations. For this reason, many have utilized phenomenological correlations to derive pseudo-redshifts of GRBs with no redshift measurement. In this work, we explore the feasibility of analytically deriving pseudo-redshifts directly from the Amati and Yonetoku relations. We simulate populations of GRBs that (i) fall perfectly on the phenomenological correlation track, and (ii) include intrinsic scatter matching observations. Our findings indicate that, in the case of the Amati relation , the mathematical formulation is ill-behaved so that it yields two solutions within a reasonable redshift range \( z \in [0.1, 10] \). When realistic scatter is included, it may result in no solution, or the redshift error range is excessively large. In the case of the Yonetoku relation, while it can result in a unique solution in most cases, the large systematic errors of the redshift calls for attention, especially when attempting to use pseudo redshifts to study GRB population properties.

\end{abstract}



\section{Introduction} \label{sec:intro}

%{\can Red: Deleted}
%{\emre Cyan: Added}

Gamma-ray bursts (GRBs) rank among the most luminous explosions in the Universe. Their isotropic and extragalactic origins were firmly established by the Burst and Transient Source Experiment (BATSE) onboard the Compton Gamma Ray Observatory \citep{meegan1992spatial}; due to their extraordinary brightness, they have been detected up to a redshift \(z = 9.4\) \citep{cucchiara2011photometric}. Thus, GRBs stand as a potential candidate to perform cosmological studies \citep{doi:10.1142/S0218271813300280}.  GRBs have traditionally been categorized by the duration of their prompt emission. Short-duration GRBs (SGRBs) are defined as those with \(T_{90} < 2\) seconds, while long-duration GRBs (LGRBs) have \(T_{90} > 2\) seconds; It is generally believed that the origin of SGRBs is attributed to the merger of binary compact objects, whereas LGRBs are associated with the deaths of massive stars \citep{zhang2009discerning, berger2014short}. However, this simple dichotomy of long versus short may not fully capture the complexity and intricacies of GRBs, since a subclass of GRBs have been found to exhibit properties of both LGRBs and SGRBs \citep{norris2006short,2023arXiv231007205Y,2024arXiv240702376W,2024arXiv241116174Y,2025arXiv250100239Z}. Moreover, using a phenomenological classification scheme based on \(T_{90}\), which depends on the detector's energy band, can also lead to contradictory results between different detectors observing the same GRB \citep{bromberg2013short}. \cite{zhang2006burst} proposes a more physical classification scheme, where GRBs are divided into Type-I (merger origin) and Type-II (collapsar origin) categories, based on their progenitor; while it may not always be possible to determine the progenitor, GRBs exhibiting characteristics of both types can be categorized based on whether the majority of its properties align more closely with a merger  or core collapse scenario.\\

\noindent In recent decades, GRBs have been found to exhibit various phenomenological correlations among parameters of the prompt emission. Among these are two robust correlations which correlate the spectral peak energy (in the \(\nu f_{\nu}\) spectrum), \(E_{\mathrm{p}}\),  to isotropic energy \(E_{\mathrm{iso}}\) \citep{2008Amati}, and isotropic peak luminosity \(L_{\mathrm{p}}\) \citep{yonetoku2004gamma}, i.e.,
\begin{equation} 
\begin{aligned}
    \log\left(\frac{E_{\text{p},z}}{\text{keV}}\right) = a_{\rm A} \log\left(\frac{E_{\text{iso}}}{\text{erg}}\right) + b_{\rm A}\,,
\end{aligned}
\label{eq:AmatiYon}
\end{equation}

and 

\begin{equation}
\log\left(\frac{E_{\text{p},z}}{\text{keV}}\right) = a_{\rm Y} \log\left(\frac{L_{\text{p},z}}{\text{erg/s}}\right) + b_{\rm Y}  \,,
\end{equation}

\begin{figure}[htbp]
    \centering
    \includegraphics[width=\linewidth]{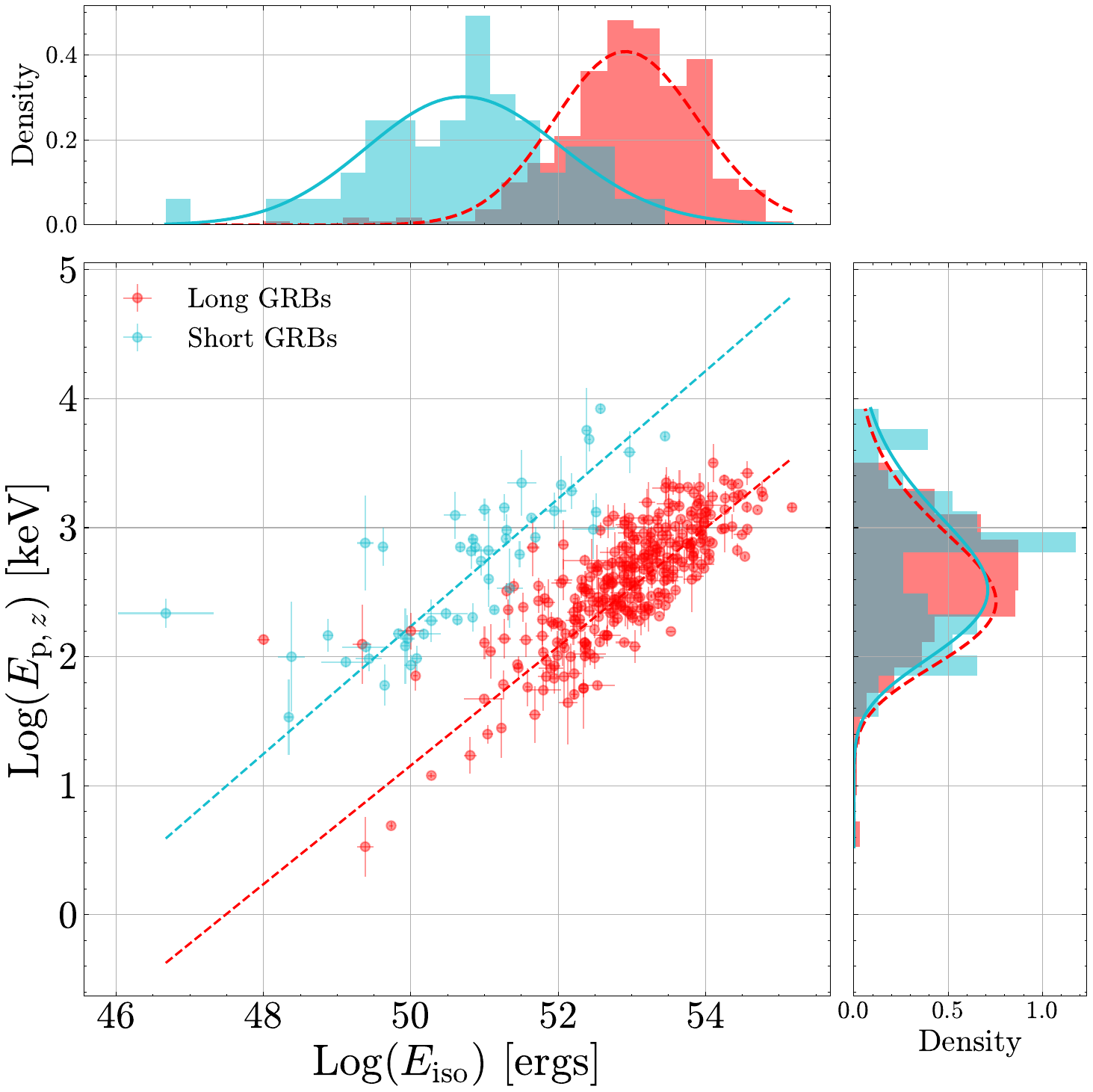}
    \caption{Fitted Amati relations. The data are taken from \cite{lan2023grb} , which includes 23 SGRBs and 333 LGRBs. The histograms on the top and bottom represent the distribution of $E_{\mathrm{iso}}$ and $E_{\mathrm{p},z}$, respectively. We fit and utilize a lognormal distribution for $E_{\mathrm{iso}}$ with a mean of $\log(E_{\mathrm{iso}})$ at $\mu = 52.9$ and $\sigma = 0.97$ for LGRBs.}
    \label{fig:amati}
\end{figure}

\noindent There exists two distinct tracks for LGRBs and SGRBs in \(E_{\mathrm{p},z}-E_{\mathrm{iso}} \) (Amati) space, with SGRBs forming a parallel track above LGRBs, due to their lower energy output (See Fig.\ref{fig:amati}).  This may be attributed to their shorter durations. However, in \(E_{\mathrm{p},z}\) -- \(L_{\mathrm{p},z}\) (Yonetoku) parameter space, there exists no significant separation between LGRBs and SGRBs. \\

\noindent The origin of these correlations remains a topic of active debate. While some have attributed them as being an artifact of instrumental selection biases \citep{Band:2005ia}, several studies have quantified the impact of selection biases and concluded that, although selection effects may influence the slope and scatter, they cannot fully account for the existence of the spectral energy correlations. Further evidence for the Amati and Yonetoku correlations comes from the discovery of an \(L - E_{\mathrm p}\) relationship that appears within individual bursts \citep{Liang:2004mr,  frontera2012broadband}. Notwithstanding the uncertainties in the underlying physics, many have considered their potential utility in deriving ``pseudo-redshifts'' \citep{yonetoku2004gamma, 2011ApJDainotti, 2013ApJTan, 2013MNRASTsutsui, deng2023pseudo}, which would be particularly vital for population studies. For example, \citet{zhangwang} applied the Yonetoku relation to constrain the luminosity function and formation rate of SGRBs, while others, such as \citet{wanderman} and \citet{paul2018modelling}, have investigated the evolution of the luminosity function in depth. Given that \textit{Fermi}-GBM has amassed a dataset of over 2,400 GRB detections since its launch \citep{poolakkil2021fermi}, we believe it is pertinent to evaluate the viability of using these correlations for deriving pseudo-redshifts.
\\

\noindent In this study, we test the feasibility of using the Amati and Yonetoku relations to analytically derive pseudo-redshifts by utilizing a synthetic catalogue of GRBs and comparing the inferred psuedo-redshifts (henceforth denoted as \(z_{\mathrm i}\)) to their true, generated redshift \(z_{\mathrm g}\). {\emre We mainly consider the case for LGRBs as they have traditionally been used for cosmological studies; however, we also provide some analysis for SGRBs in the Discussion.} The paper is organized as follows: section \ref{methods} describes our detailed simulation process; in section \ref{results}, we summarize our results, and section \ref{summary} offers a summary and discussion.

%We simulate 100 LGRBs with intrinsic paramaters of \( E_{\mathrm{p,z}}  \), \(E_{\mathrm{iso}}\) and \( z_{\mathrm g} \).  For each GRB, we are thus able to obtain the observed fluence \(F\) and observed spectral peak energy, \(E_{\mathrm{p,o}}\). 

\section{Methods} \label{sec:style}
\label{methods}

 \noindent {\emre 
The Amati correlation is defined in terms of two \emph{intrinsic} (i.e., rest-frame) quantities: the peak energy \(E_{\mathrm{p},z}\) and the isotropic energy \(E_{\mathrm{iso}}\), for a given redshift \(z\).} The intrinsic parameters \(E_{\mathrm{p},z}\) and \(E_{\mathrm{iso}}\) are related to their observed counterparts through 
\begin{equation} 
E_{\text{p},z} = E_{\text{p,o}} \times (1 + z)\,  
\label{eq:intrinsicEp}
\end{equation}
and 
\begin{equation}
E_{\mathrm{iso}} = \frac{4\pi D^2_{\mathrm{L}} F k}{1 + z} \,,
\label{eq:intrinsicEiso}
\end{equation}
where \(F\) is the fluence, \(k\) is the k-correction (which we take as \(k = 1\), due to the wide energy coverage of \textit{Fermi}-GBM), and \(D_{\mathrm{L}}\) is the luminosity distance, i.e., 
\begin{equation}
D_L(z) = (1 + z) \frac{c}{H_0} \int_0^z \frac{dz'}{\sqrt{\Omega_M (1 + z')^3 + \Omega_\Lambda}}\,.
\end{equation}

 \noindent We adopt the Planck18 cosmology, assuming a flat $\Lambda$CDM universe with $H_0 = 67.66 \, \mathrm{km} \, \mathrm{s}^{-1} \, \mathrm{Mpc}^{-1}$ and $\Omega_{\mathrm{M}} = 0.3111$ as reported by \cite{aghanim2020planck}. { \emre Holding \(E_{\mathrm{p,o}}\) and \(F\) fixed for a given burst, and allowing \(z\) to vary continuously (over \(z\in [0.1,10]\)), we trace out a continuous curve in the \(E_{\mathrm{iso}}- E_{\mathrm{p},z}\) plane.  We refer to this locus of points as 

\begin{equation}
\mathcal{A}(z; E_{\mathrm{p,o}}, F) = \left( E_{\mathrm{iso}}(z),\; E_{\mathrm{p},z}(z) \right).
\end{equation}

\noindent Geometrically, it is a parametric curve where the running parameter is \(z\). We next compare \(\mathcal{A}(z; E_{\mathrm{p,o}}, F)\) to the best-fit Amati line;
any \emph{intersection} between \(\mathcal{A}(z; E_{\mathrm{p,o}}, F)\) and this correlation line corresponds to a candidate “pseudo-redshift” \(z_i\): that is, a redshift for which the observed \((F, E_{\mathrm{p,o}})\) values place the burst exactly on the Amati relation. 

\noindent We fit the Amati parameters \((a_{\rm A}, b_{\rm A})\) to the LGRB sample of \citet{lan2023grb}, finding \(a_{\rm A} = 0.46\pm0.016\) and \(b_{\rm A}=-21.84\pm0.87.\)  We first simulate a population of 100 GRBs that perfectly obey this best-fit Amati line \footnote{All the relevant codes in this paper can be accessed from the \texttt{GitLab} repository \href{https://code.ihep.ac.cn/emre/pseudo-redshifts}{\texttt{</>}}
 }.  Specifically, we first draw \(E_{\mathrm{iso}}\) values from the observed distribution of \citet{lan2023grb} and sample the “true” redshift \(z_{\mathrm{g}}\) for each GRB from a typical LGRB formation rate \(\psi(z)\) \citep{salvaterra2007gamma}:

\begin{equation}
     \Psi_{\rm GRB}(z) = k_{\rm GRB} \, \Sigma (z) \, \psi_*(z)
\end{equation}

\noindent where $\Sigma (z)$ is the metallicity-convolved efficiency function \citep{kewley2005metallicity}

\begin{equation}
    \Sigma(z) = \frac{\Gamma\left[0.84, \left(\frac{Z_{\text{th}}}{Z_{\odot}}\right)^{2} 10^{0.3z}\right]}{\Gamma(0.84)}
\end{equation}

\noindent $\psi_*(z)$ is the star formation rate from \cite{li2008star} and we take the normalization factor $k_{\rm GRB} = 1.0$. For each simulated burst, we set 
\[
E_{\mathrm{p},z} \;=\;10^{\,\{a_{\rm A}\,\log_{10}(E_{\mathrm{iso}}) + b_{\rm A}\}},
\]
and then compute its observed fluence and observed peak energy via Equations. \ref{eq:intrinsicEp} and \ref{eq:intrinsicEiso}. 

\noindent Thus, each simulated GRB is assigned both intrinsic parameters \(\{E_{\mathrm{iso}},\,E_{\mathrm{p},z_{\rm g}},\,z_{\mathrm{g}}\}\) and observed parameters \(\{F,\,E_{\mathrm{p,o}}\}\).  Finally, to \emph{infer} its pseudo-redshift, we construct the parametric curve \(\mathcal{A}(z; E_{\mathrm{p,o}}, F)\) and find any intersection(s) with the Amati line.  The resulting solution(s) \(z_i\) can then be compared to the known “true” \(z_{\mathrm{g}}\). Reliable distance indicators must yield only a single redshift solution within a reasonable redshift range, and so we wish to see if the Amati relation can yield a unique solution \(z_i\) for each GRB, given its observed spectral peak energy \(E_{\mathrm{p,o}}\) and fluence \(F\).   Next, we add intrinsic scatter to the simulated data, based on our measured dispersion of \(\sigma_{ \log E_{{\rm p}, z}}\) = 0.30 on the data of \cite{lan2023grb}, where we attempt to gauge the statistics of $z_{\rm i}$ with a simulated catalogue of 500 GRBs; this dispersion defines an upper and lower boundary around the best-fit Amati relation, forming the uncertainty region \(\mathcal{U}_{\mathrm A}\). For each simulated GRB, we determine its 1\(\sigma\) redshift range \(\bigl[z_{\mathrm i}^{-\sigma},\, z_{\mathrm i}^{+\sigma}\bigr]\) by finding where the parametric curve \(\mathcal{A}(z; E_{\mathrm{p,o}}, F)\) intersects \(\mathcal{U}_{\mathrm A}\).   }\\

\begin{figure*}[t]
  \centering
  \begin{subfigure}[t]{0.491\textwidth} % Use nearly half the text width for each subfigure
    \centering
    \includegraphics[width=1.04\textwidth]{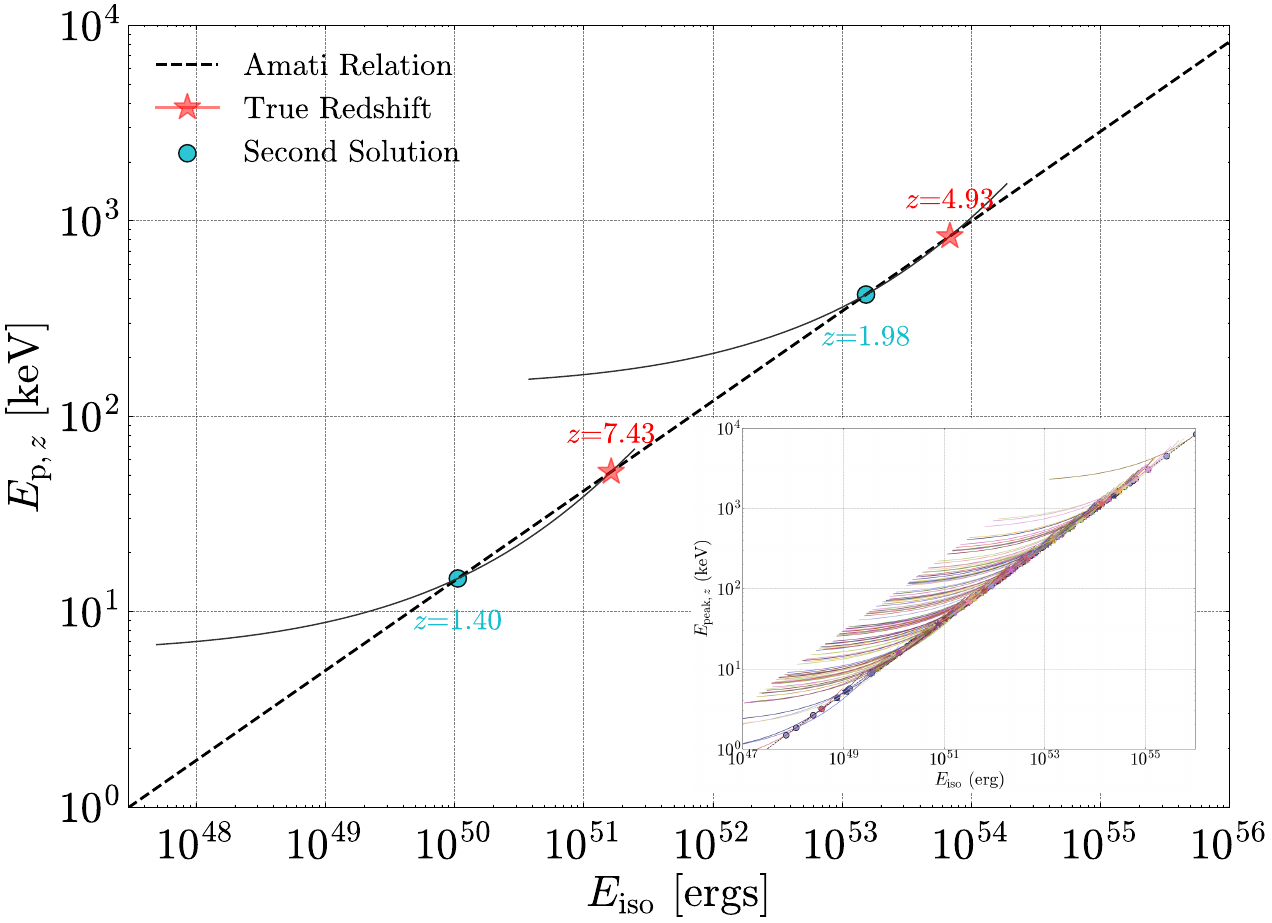} 
  \end{subfigure}
  \hspace{0.0001\textwidth}
  \begin{subfigure}[t]{0.49\textwidth} % Adjust width to fit perfectly next to each other
    \centering
    \includegraphics[width=1.01\textwidth]{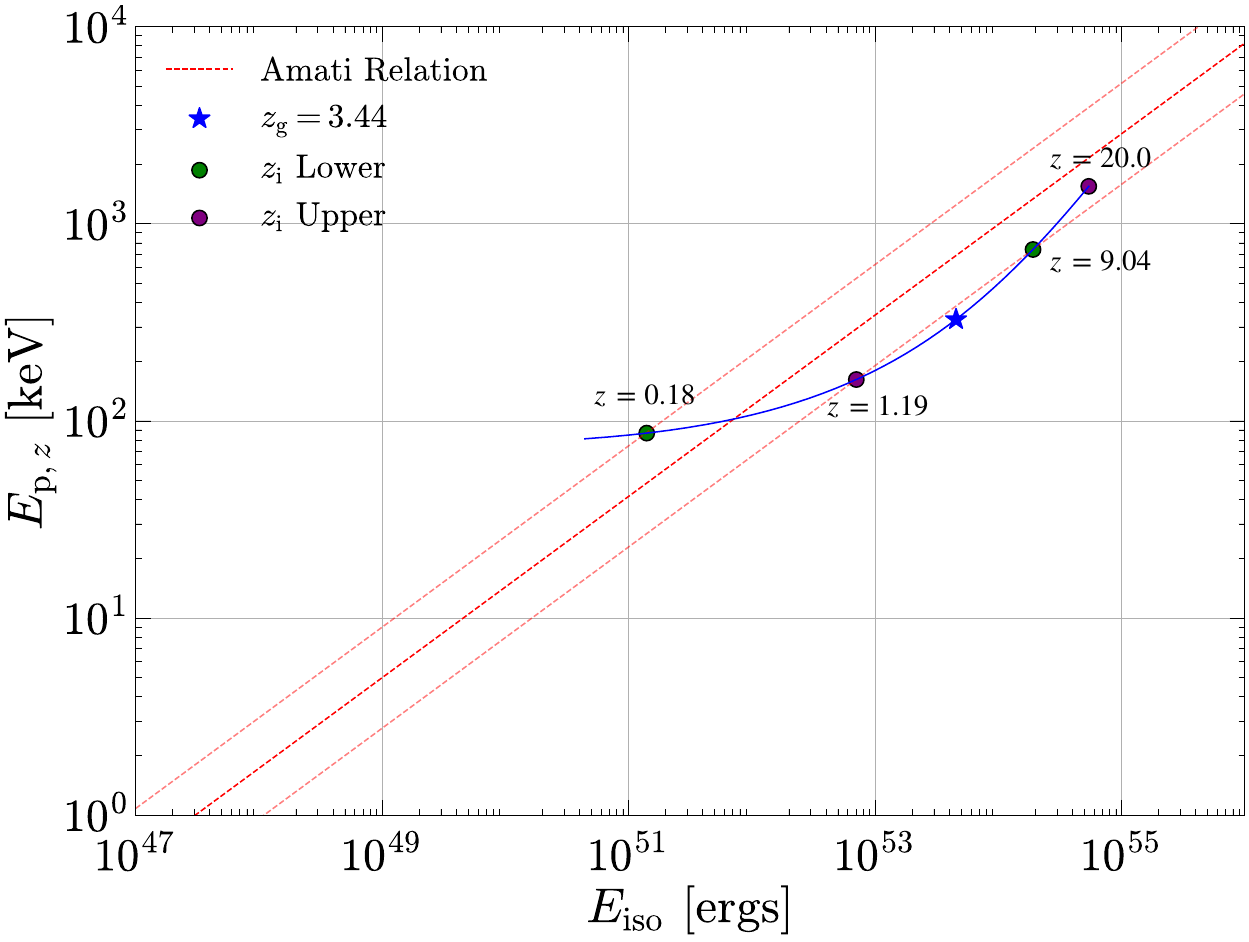}
  \end{subfigure}

  \vspace{0.02cm} % Reduce the space between the rows
 \centering
  \begin{subfigure}[t]{0.49\textwidth}
    \centering
    \includegraphics[width=1.02\textwidth]{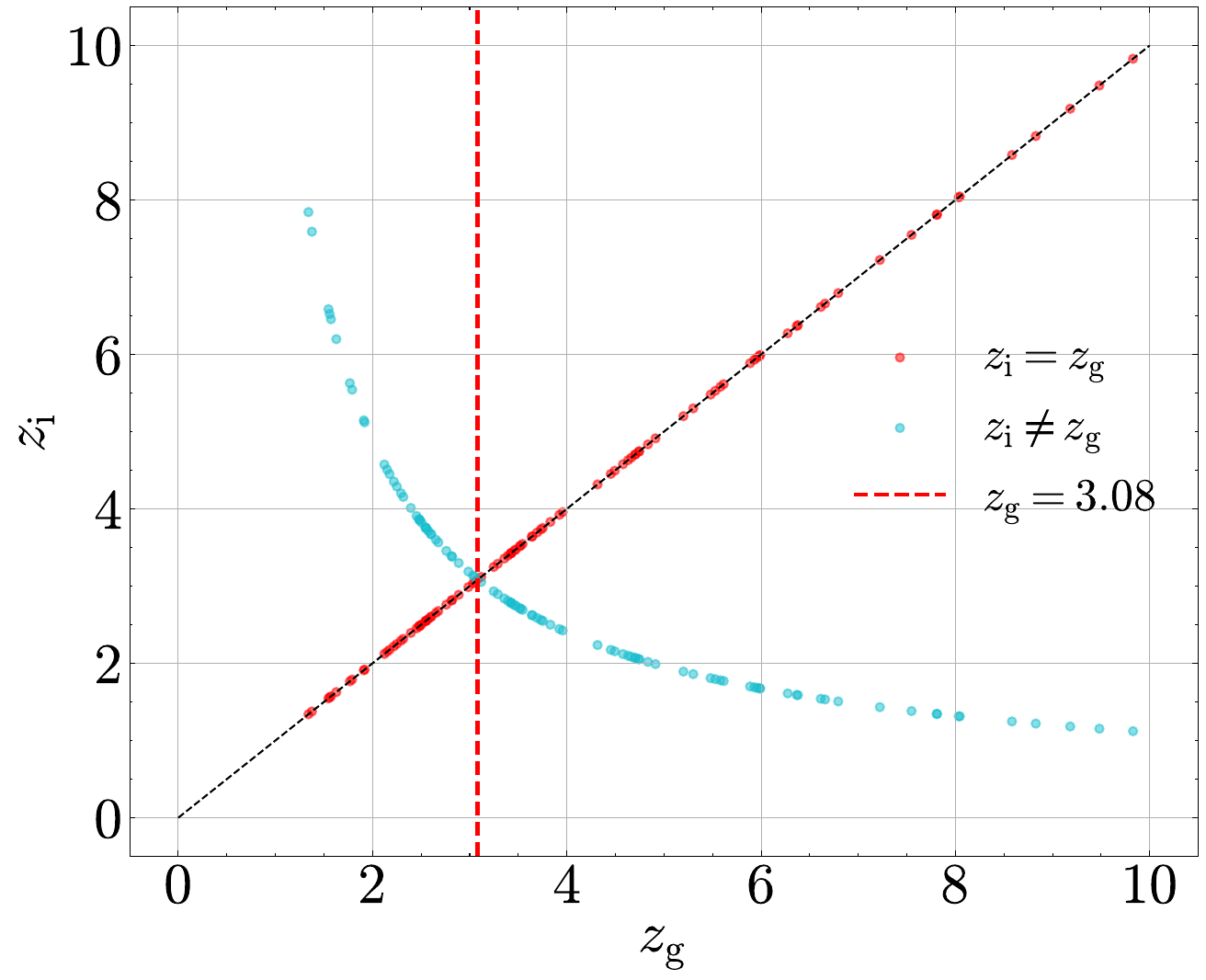} 
  \end{subfigure}
  \hspace{0.001\textwidth}
  \centering
  \begin{subfigure}[t]{0.49\textwidth}
    \centering
    \includegraphics[width=1.02\textwidth]{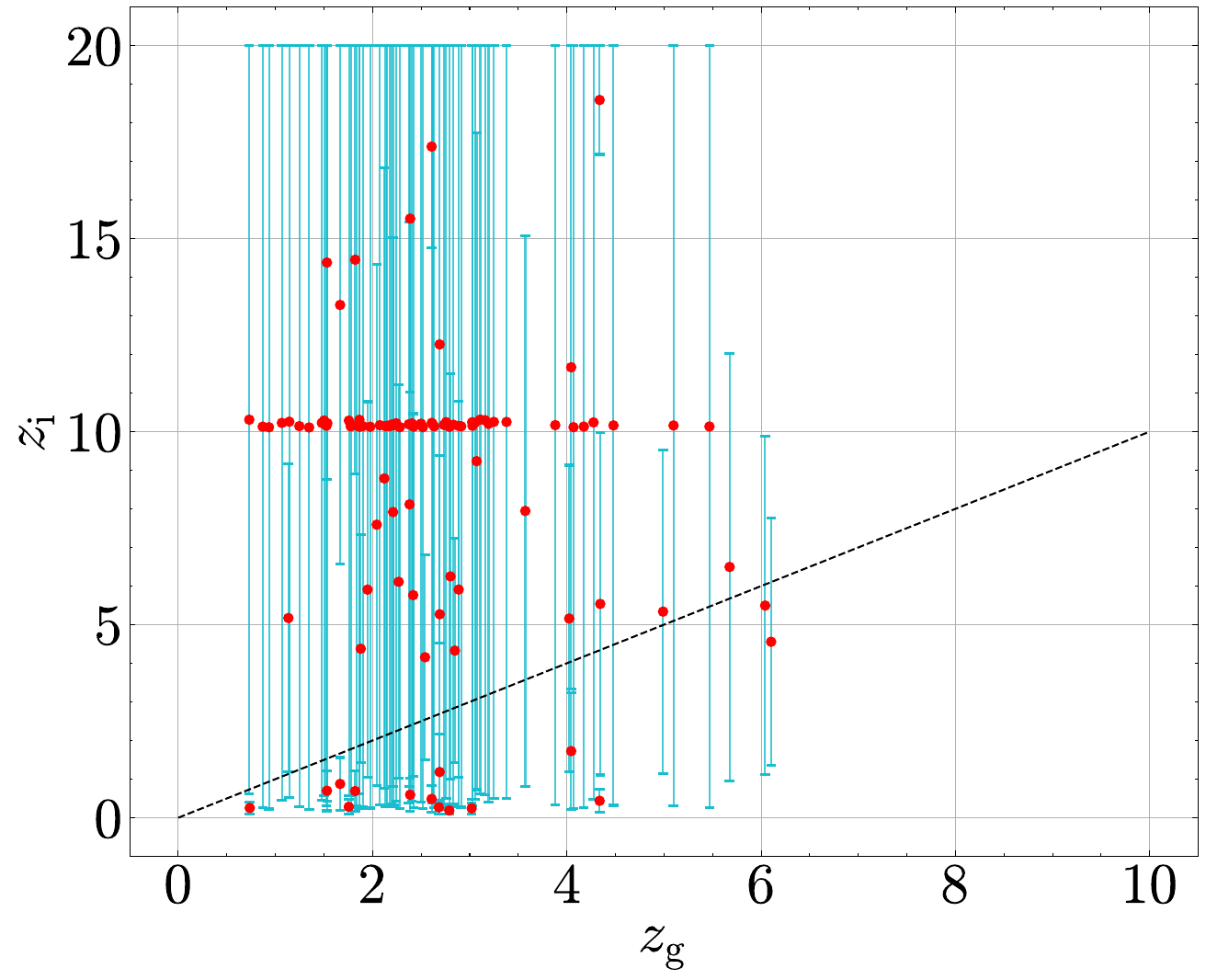}
  \end{subfigure}

  \caption{\textbf{Top Left}:  Illustration of two simulated \( \mathcal{A}(z; E_{\mathrm{p,o}}, F) \)  curves; many more examples are shown in the inset. Red stars denote the true positions and \(z_{\mathrm g}\) in Amati parameter space; blue circle denotes the second inferred position and redshift \(z_{\mathrm i}\). \textbf{Bottom Left:} \(z_{\mathrm i}\) vs \(z_{\mathrm g}\); the vertical red dashed line denotes the \(z_{\mathrm t}\) prediction from eq. \ref{eq:limit_amati}. Clearly,  \(z_{\mathrm g} = z_{\mathrm i}\) and  \(z_{\mathrm g} \neq z_{\mathrm i}\) {\emre solution branches} intersect at \(z = 3.08 \) as predicted \href{https://code.ihep.ac.cn/emre/pseudo-redshifts/-/blob/main/Amati.py?ref\_type=heads}{\texttt{</>}} \textbf{Top Right:} A sample curve with two crossings when scatter in the Amati relation is included. The real position of the GRB is denoted by the blue star, with \(z_{\mathrm g} = 3.44\). \textbf{Bottom Right:} Confidence intervals \([z_{\mathrm i}^{-\sigma}, z_{\mathrm i}^{+\sigma}]\) of 100 GRBs obtained from the intersection of the 1\(\sigma\) bounds of the Amati relation, \(\mathcal{U_{\mathrm A}}\).  We plot the confidence intervals of both GRBs with single crossings and double crossings. The red dot denotes the midpoint redshift within each band. \href{https://code.ihep.ac.cn/emre/pseudo-redshifts/-/blob/main/Amati-Errors.py}{\texttt{</>}}
 }
  \label{fig:perfectamati}
\end{figure*}

\noindent  In the case of the Yonetoku relation, the intrinsic luminosity is given by  
 \begin{equation}
L_{z} = 4\pi D^2_{\rm L} f_{\gamma} k\,,
\end{equation}
where \(f_{\gamma}\) is the observed flux. {\emre Since we are substituting $E_{\rm iso}$ with luminosity, } the GRB parametric curve \(\mathcal{Y}( z; E_{\mathrm{p,o}} , f_{\gamma} )\) in Yonetoku parameter space takes on a distinct functional form with respect to \( z \) than \(\mathcal{A}(z; E_{\mathrm{p,o}}, F)\); in this case, each simulated GRB in Yonetoku parameter space is characterized by the parameters \( \{ L_{\mathrm{p}}, E_{\mathrm{p},z_{\mathrm g}}, f_{\gamma}, E_{\mathrm{p,o}}, z_{\mathrm g} \} \) and warrants a separate investigation. We follow a similar procedure outlined for the Amati relation when assigning the intrinisic parameters of the simulated GRBs, and {\emre dispersion} according to \(\sigma_{ \log E_{{\rm p}, z}}\) = 0.25 \citep{ghirlanda2005peak}, which defines the Yonetoku uncertainty band \(\mathcal{U_{\mathrm Y}}\). {\emre We take \(a_{Y}\) to be 0.625 $\pm 0.032$, and \(b_{Y}\) to be -30.22 $\pm 0.023$  taken from literature  \citep{yonetoku2010possible} \footnote{From \cite{yonetoku2010possible}, \[
L_p = 10^{52.43 \pm 0.037} \times \left[ \frac{E_p (1 + z)}{355 \, \text{keV}} \right]^{1.60 \pm 0.082}
\] Thus, \(a_{Y}\) = $\frac{1}{1.6}$, and \(b_{Y}\) = $\log(355) - \frac{52.43}{1.6} $  }. We sample the luminosities from a simple lognormal distribution \( \log_{10} L_{\text{peak}} \sim \mathcal{N}(\mu = 52.5, \sigma^2 = 1) \). {\emre Note that the choice of luminosity function does not alter the key results, because the functional form of \(\mathcal{Y}(z; E_{\mathrm{p,o}}, f_{\gamma})\)—and hence the geometry of its intersection with the Yonetoku correlation—remains unchanged. Consequently, only the relative weighting of the parameter space is affected, not the overall behavior of  solutions.    }} \\

\section{Results}
\label{results}
\subsection{Amati Relation}
%\label{results}
\noindent In the top left panel of Figure \ref{fig:perfectamati}, we display two example curves of GRBs simulated without scatter (assuming perfect Amati relation) overlaid on the best-fit Amati line within the Amati parameter space. It is evident that these GRBs 
have double intersections with the Amati line. In fact, unless the parametric curve is exactly tangent to the Amati line—which occurs only at a specific redshift \( z_{\mathrm t} \)—there will be two redshift solutions for each GRB, {\emre one with $z_{\rm i} < z_{\mathrm t}$ and another with $z_{\rm i} > z_{\mathrm t}$} . We derive an expression for \( z_{\mathrm t}\), which is dependant on the Amati slope \(a_{\mathrm A}\) (see Appendix):
\begin{equation}
    a_{\mathrm A} = \frac{1}{2 \dfrac{d}{d \log(1+z_{\mathrm t})} \left( \log D_{\rm L}(z_{\mathrm t}) \right) - 1}
    \label{eq:limit_amati}\,.
\end{equation}

\noindent  In the bottom left panel of Fig. \ref{fig:perfectamati}, we show that analytically solving the Amati relation yields two solutions, with (i)  \(z_{\mathrm i} = z_{\mathrm g}\) and (ii) \(z_{\mathrm i} \neq z_{\mathrm g}\); {\emre  the two solution branches} of \(z_{\mathrm i} = z_{\mathrm g}\) and \(z_{\mathrm i} \neq z_{\mathrm g}\) intersect at \(z_{\rm t} = 3.08\), as predicted by Eq \ref{eq:limit_amati} (See left panel of Figure \ref{fig:limitingslope} in the appendix).  {\emre The condition that determines whether the Amati relation is ``ill-behaved'' in the physical redshift interval \(z \in [0.1,10]\) depends on how steep \(a_{\mathrm A}\) is. Specifically, one can require \(z_{\mathrm t}>10\) ensuring every parametric curve to admit a single solution rather than two. This translates to a limiting slope \(a_{\mathrm A}\geq 0.67\), significantly larger than the best-fit value currently measured (see the left panel of Fig.~\ref{fig:limitingslope}) in the appendix} \\  

\noindent When we incorporate intrinsic scatter into the simulated GRBs as previously described,  the \( \mathcal{A}(z; E_{\mathrm{p,o}}, F)\) curves will either have no solution, a single solution, or double solution. As a direct corollary of Eq \ref{eq:limit_amati}, we see that taking the redshift at which    \( \mathcal{A}(z; E_{\mathrm{p,o}}, F) \) is closest to the mean Amati line for those curves which do not intersect the Amati line will always result in the same redshift, \(z_{\mathrm t}\). The  \( \mathcal{A}(z; E_{\mathrm{p,o}}, F) \) curves which yield only a single solution (about 34\% of the sample) will always have \(z_{\mathrm i} < z_{\mathrm t}\) (see Fig \ref{fig:onesolution} in appendix). \\

\noindent When we consider the intersection of \( \mathcal{A}(z; E_{\mathrm{p,o}}, F) \) with the Amati {\emre dispersion} area \( \mathcal{U_{\mathrm A}} \), the curves would either have no crossings {\emre (10\%)} , single crossings, or double crossings {\emre (8\%)} . In the bottom right panel of Fig. \ref{fig:perfectamati}, we plot the confidence intervals \([z_{\mathrm i}^{-\sigma}, z_{\mathrm i}^{+\sigma}]\) against \(z_{\mathrm g}\), including both GRBs with single crossings or double crossings.  Clearly, even with the  1\(\sigma\) uncertainty band, the uncertainty range of \(z_{\mathrm i}\) is around the same order of magnitude as the redshift range of interest, and in fact often extends beyond \(z_{\mathrm i} > 30\). There exists no correlation between \(z_{\mathrm i}\) and \(z_{\mathrm g}\). Note that the case of single crossings here is intrinsically different from the case of  \( \mathcal{A}(z; E_{\mathrm{p,o}}, F)\) with a single intersection (at \(z_{\mathrm t}\)) when simulated without scatter. 

\begin{figure}[h!]
    \centering
    \includegraphics[width=\linewidth]{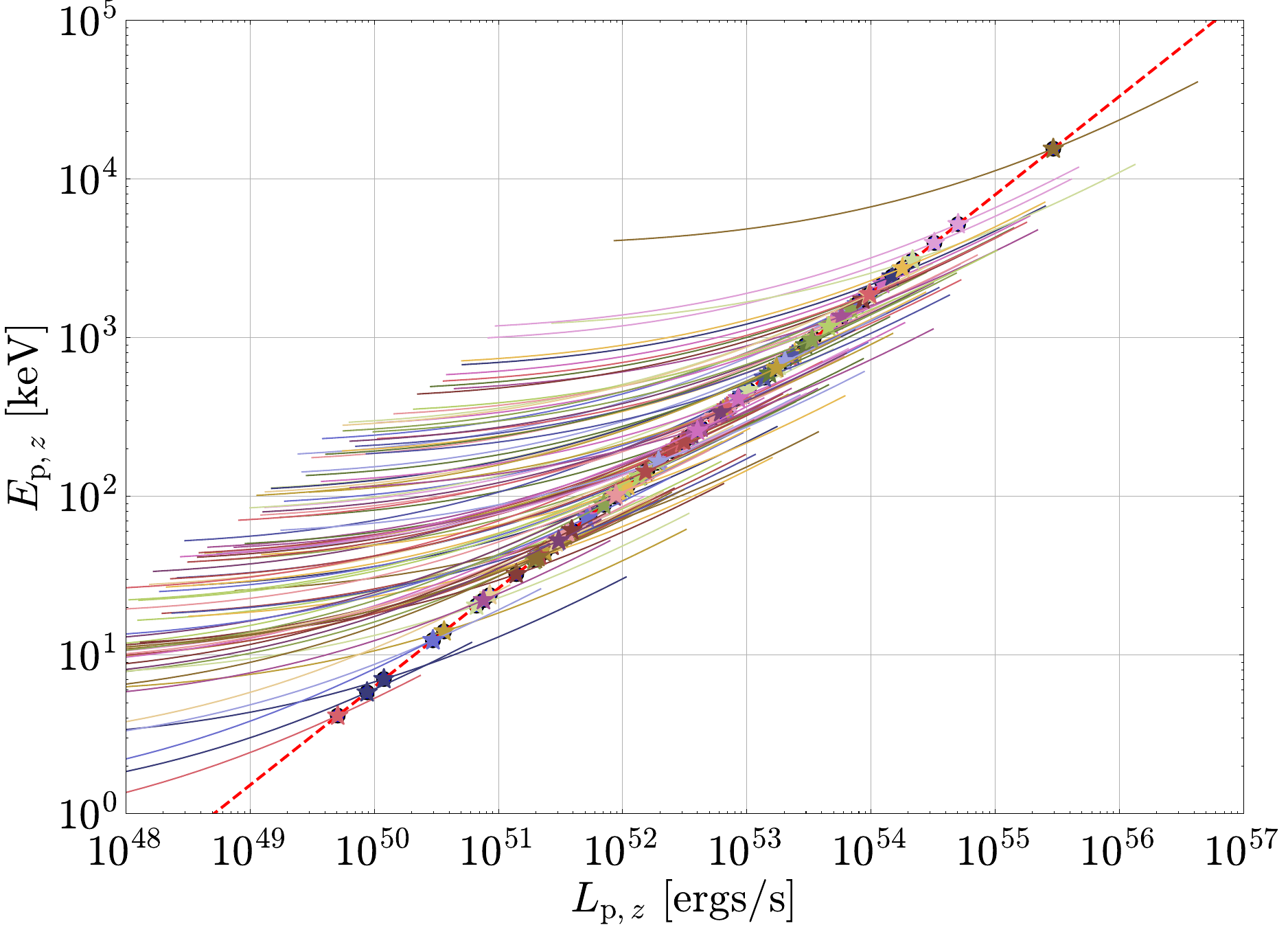}
    \caption{Plot lot of 100  \(\mathcal{Y}( z; E_{\mathrm{p,o}} , f_{\gamma} )\) curves of GRBs simulated perfectly on the Yonetoku track. The functional form of \(\mathcal{Y}( z; E_{\mathrm{p,o}} , f_{\gamma} )\) is clearly distinct from  \( \mathcal{A}(z; E_{\mathrm{p,o}}, F) \), resulting in one intersection for \(z \in [0.1, 10] \href{https://code.ihep.ac.cn/emre/pseudo-redshifts/-/blob/main/Yonetoku-NoScatter.py}{\texttt{</>}}
  \)}
    \label{fig:yonetoku-1}
\end{figure}

\begin{figure*}[t]
  \centering
  \begin{subfigure}[t]{0.49\textwidth} % Use nearly half the text width for each subfigure
    \centering
    \includegraphics[width=1.02\textwidth]{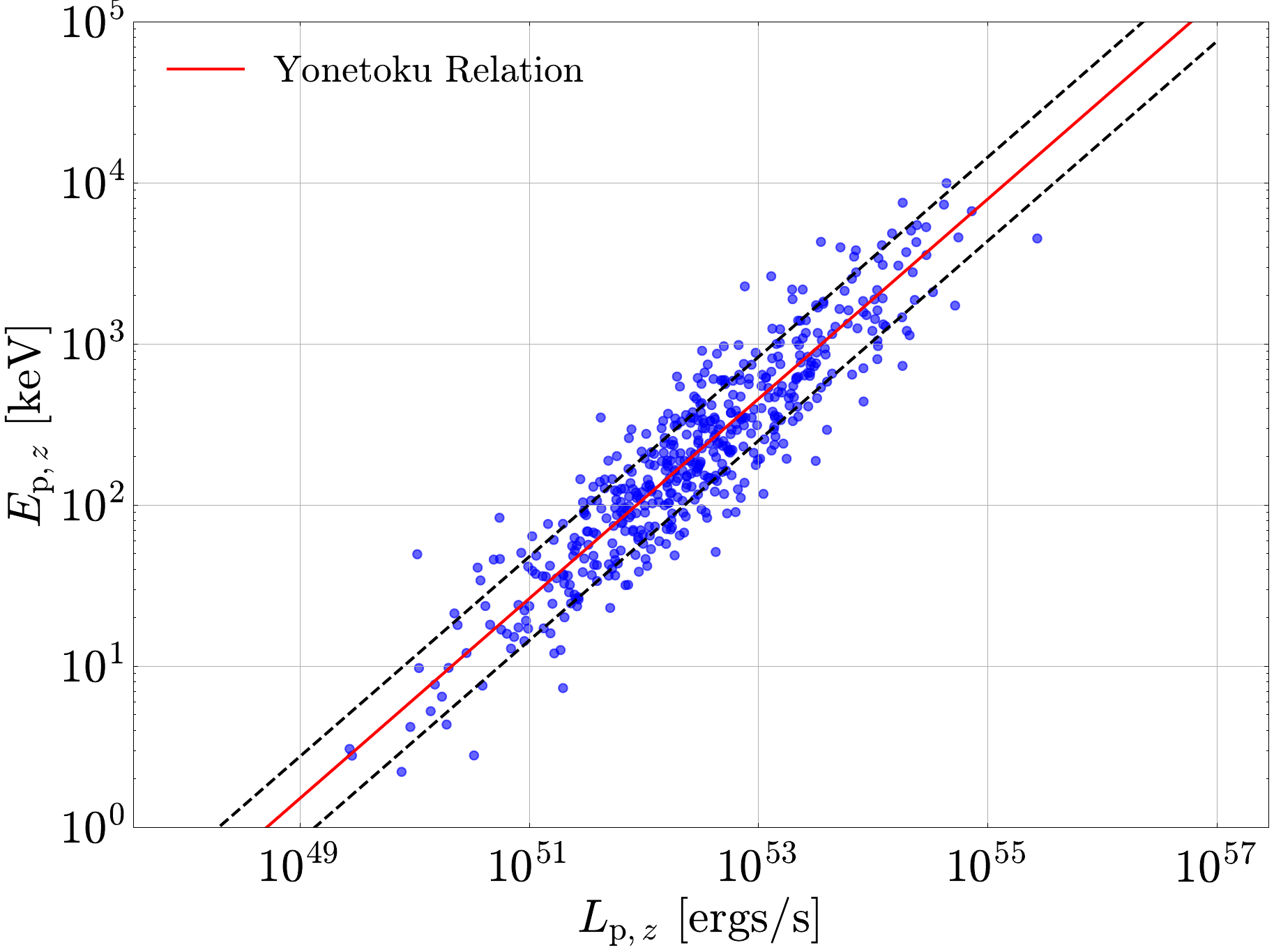} 
  \end{subfigure}
  \hspace{0.001\textwidth}
  \begin{subfigure}[t]{0.49\textwidth} % Adjust width to fit perfectly next to each other
    \centering
    \includegraphics[width=1.05\textwidth]{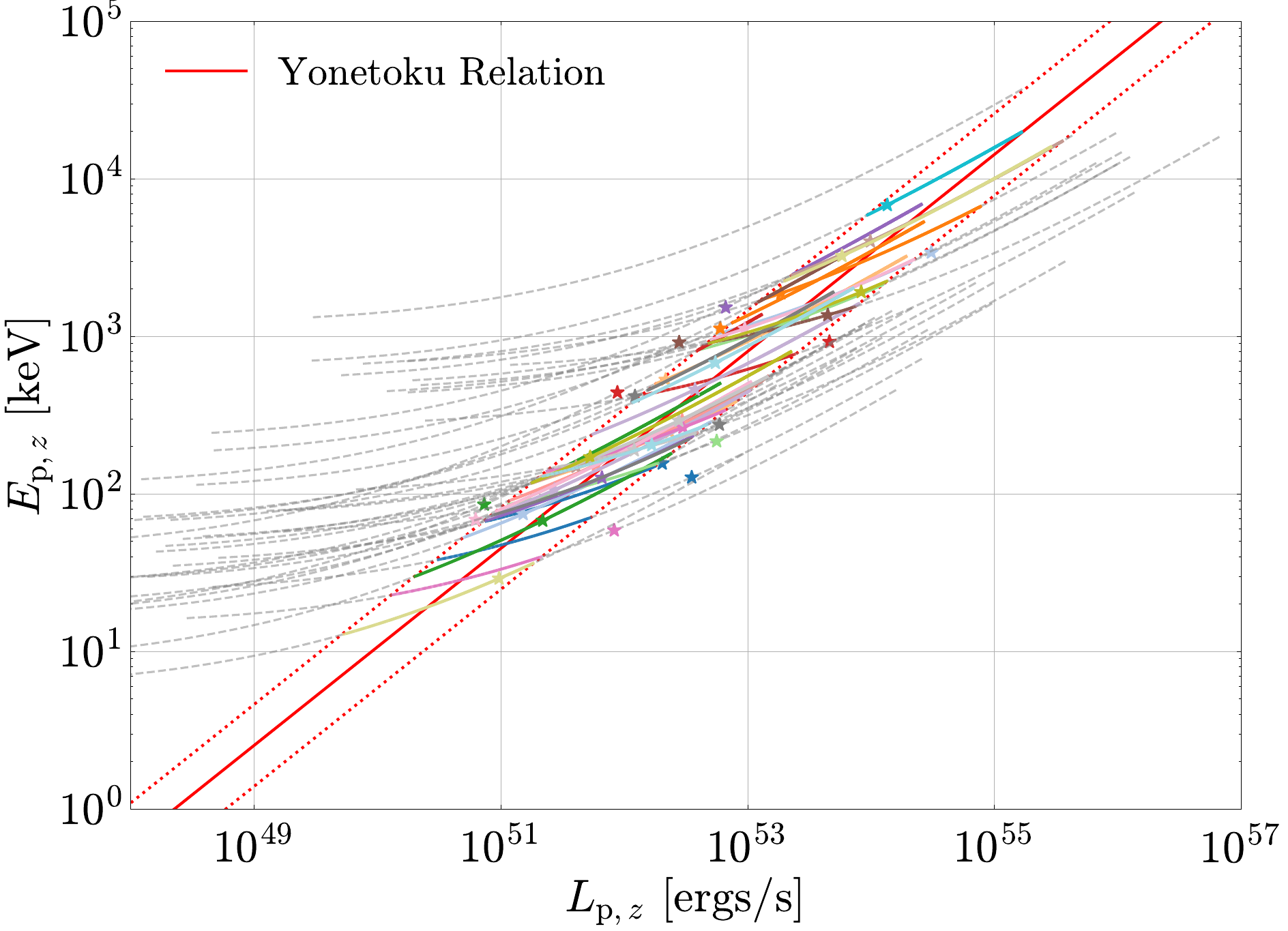}
  \end{subfigure}

  \vspace{0.05cm} % Reduce the space between the rows
 \centering
  \begin{subfigure}[t]{0.49\textwidth}
    \centering
    \includegraphics[width=1.03\textwidth]{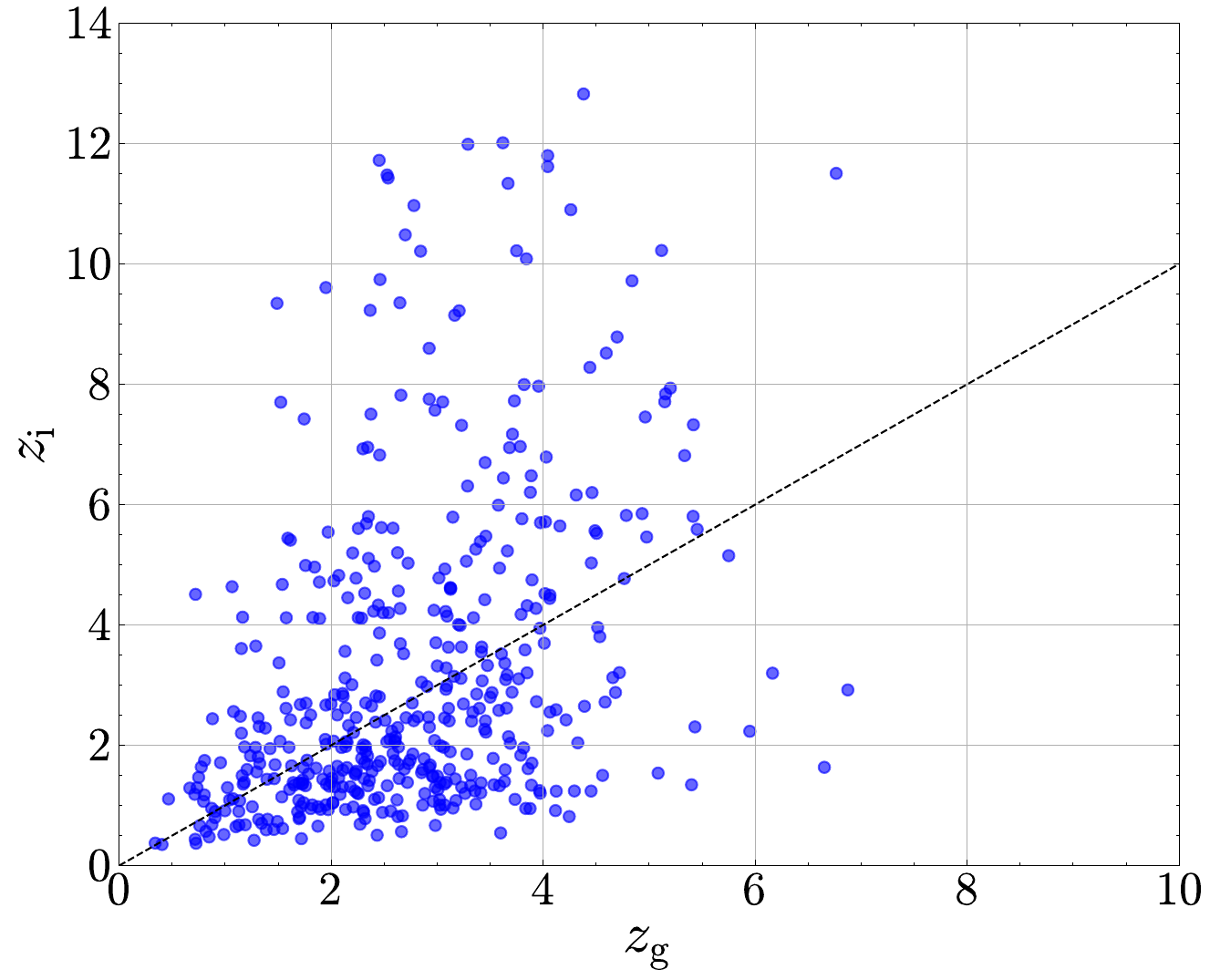} 
  \end{subfigure}
  \hspace{0.001\textwidth}
  \centering
  \begin{subfigure}[t]{0.49\textwidth}
    \centering
    \includegraphics[width=1.06\textwidth]{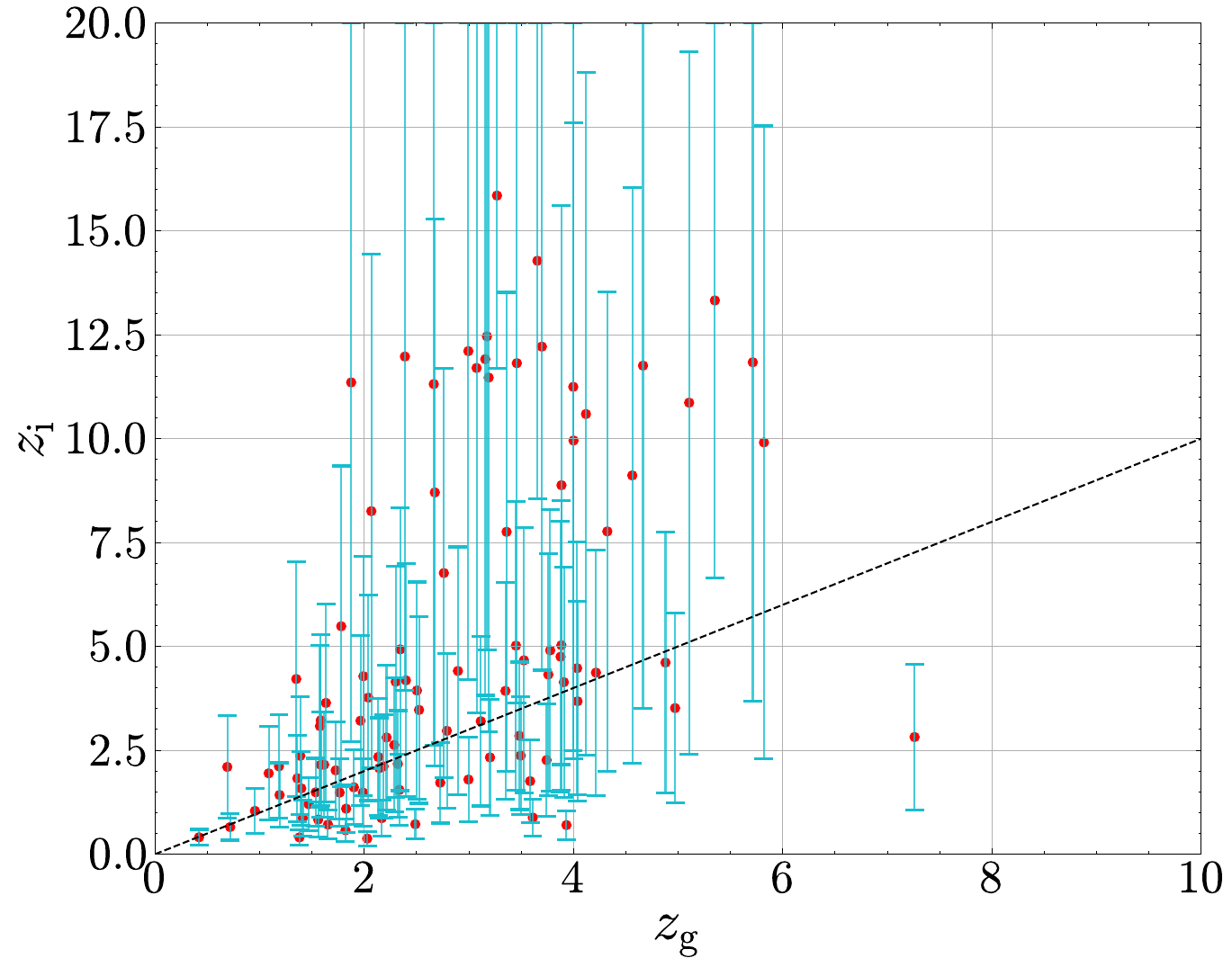}
  \end{subfigure}

  \caption{\textbf{Top Left:} Simulated GRBs in Yonetoku parameter space with intrinsic scatter \(\sigma_{ \log E_{\rm p}, z} = 0.25 \) . \textbf{Bottom Left:} \(z_{\mathrm i}\) vs \(z_{\mathrm g}\), with a Pearson Correlation Coefficient of $r = 0.40 \pm 0.04$ \href{https://code.ihep.ac.cn/emre/pseudo-redshifts/-/blob/main/Yonetoku\%20-\%20Scatter?ref\_type=heads}{\texttt{</>}}. \textbf{Top Right:} 40 Simulated \(\mathcal{Y}( z; E_{\mathrm{p,o}} , f_{\gamma} )\) curves (for \( z \in [0.1, 30] \)) over the 1 \(\sigma \) Yonetoku uncertainty area \( \mathcal{U_{\mathrm Y}}\). The length of \(\mathcal{Y}( z; E_{\mathrm{p,o}} , f_{\gamma} )\) within \( \mathcal{U_{\mathrm Y}}\) is highlighted in color, and the stars denote the true GRB position.  \textbf{Bottom Right:} Confidence intervals \([z_{\mathrm i}^{-\sigma}, z_{\mathrm i}^{+\sigma}]\) vs \(z_{\mathrm g}\) of  {\emre 100 } simulated GRBs derived from the intersection of  \(\mathcal{Y}( z; E_{\mathrm{p,o}} , f_{\gamma} )\) with the Yonetoku uncertainty area \( \mathcal{U_{\mathrm Y}} \) \href{https://code.ihep.ac.cn/emre/pseudo-redshifts/-/blob/main/Yonetoku-Errors.py?ref\_type=heads}{\texttt{</>}}
}
  \label{fig:yonetoku2}
\end{figure*}

\subsection{Yonetoku Relation}
\label{Yonetokuresults}

\noindent  {\emre Because \(L_{\mathrm{p},z}\) is more sensitive to redshift than \(E_{\mathrm{iso}}\), the parametric curve \(\mathcal{Y}(z; E_{\mathrm{p,o}}, f_{\gamma})\) does not intersect the correlation line multiple times within \(z \in [0.1,10]\). As a result, each burst yields a \emph{unique} solution for \(z_{\mathrm i}\) (see Fig.~\ref{fig:yonetoku-1}). }. With the same line of reasoning adopted for the case of the Amati relation, we can obtain a threshold slope for which the Yonetoku relation would no longer always yield a unique solution within \( z \in [0.1, 10] \), through 
\begin{equation}
    a_{\mathrm Y} = \frac{1}{2 \dfrac{d}{d \log(1+z_{\mathrm t})} \left( \log D_{\rm L}(z_{\mathrm t}) \right) }\,,
\end{equation}

\noindent where we find that the condition of being ill-behaved is satisfied for \(a_{\mathrm Y} \leq 0.41\), well below the uncertainty range for \(a_{\mathrm Y}\) (See right panel of Fig \ref{fig:limitingslope} in Appendix). Therefore, the Yonetoku relation is always well behaved and provides a unique redshift solution within \( z \in [0.1, 10] \). When intrinsic scatter is added to \(E_{{\rm p}, z}\), we find that about {\emre 5\%} of GRBs do not intersect the Yonetoku line. In the bottom left panel of Fig. \ref{fig:yonetoku2}, we plot  \(z_{\mathrm i}\) vs \(z_{\mathrm g}\). It is evident that the \(z_{\mathrm i}\) vs \(z_{\mathrm g}\) trend is highly sensitive to the intrinsic scatter of the sample; we obtain a pearson correlation coefficient of {\emre $r = 0.40 \pm 0.04$} , indicating little predictive power. \\

\noindent In the bottom right panel of Figure \ref{fig:yonetoku2}, we show \([z_{\mathrm i}^{-\sigma}, z_{\mathrm i}^{+\sigma}]\) vs \(z_{\mathrm g}\), found from the intersection of  \(\mathcal{Y}( z; E_{\mathrm{p,o}} , f_{\gamma} )\) with the yonetoku dispersion area \( \mathcal{U_{\mathrm Y}} \); unlike in the case of the Amati relation, we do not find any curves with double crossings, and a small fraction (\(1\%\)) of curves have no crossings. Nonetheless, the confidence bands are excessively large, often extending well beyond \( z_{\mathrm i} > 30\). We cap \(z_{\mathrm i}\) to 20.

\section{Summary \& Discussion }
\label{summary}
\noindent In this paper, we examine the feasibility of analytically obtaining pseudo-redshifts using two well-known phenomenological correlations of the prompt emission, which have become the de facto method for this purpose. With the aid of a synthetic catalogue of GRBs, we find this practice to be untenable for the following reasons: \begin{itemize}
    \item Analytically solving for \(z_{\rm i}\) from the best fit Amati relation always results in two solutions, which are both within a physical range (except in the case of \(z_{\mathrm g} = z_{\mathrm t}\)). The Amati relation can be practically well behaved over \( z \in [0.1, 10] \) if the Amati slope \(a_{\mathrm A} \geq 0.67 \), well above the uncertainty range of \(a_{\mathrm A}\). Intrinsic scatter would not salvage the situation; if intrinsic scatter is added, there would either be (i) two solutions, (ii) one solution, or (iii) no solution. In the case of no solution, taking the point on  \( \mathcal{A}(z; E_{\mathrm{p,o}}, F) \) closest to the mean Amati track would only yield a constant \(z_{\mathrm i}\), as a corollary of Equation \ref{eq:limit_amati}. In the case of a single solution, it is easy to see that in all cases \(z_{\mathrm i} < z_{\mathrm t}\), and hovers around a constant value. {\emre Therefore, the Amati relation fails at a fundamental level and cannot be reliably used as a distance indicator}
    \item Although the Yonetoku relation results in a unique solution for \(z_{\mathrm{i}}\), the intrinsic scatter leads to there being no solution for about {\emre 5\%} simulated GRBs and little predictive power of \(z_{\mathrm i}\).  
    \item Incorporating the confidence areas \(\mathcal{U_{\rm A}}\) and \(\mathcal{U_{\rm Y}}\)  would yield excessively large error bands for \(z_{\mathrm i}\) for both the Amati and Yonetoku relations. 
\end{itemize}
\noindent  
We have not simulated GRBs with uncertainty bands in \(E_{\mathrm p}\) and \(L_{\mathrm p}\), so in practice, we would expect even larger uncertainties in \(z_{\mathrm i}\) if the uncertainties of fluence and peak energy are accounted for. {\emre In order to check whether the redshift distribution of the simulated GRB population will change the above analysis on the Yonetoku relation, we repeat the analysis with a different redshift distribution, a density evolved rate, as proposed in Lan et al, ie,   $\psi_{\rm GRB} = \psi_{*}(z) (1 + z)^{\delta}$, with $\delta = 1.9$, which places more GRBs at higher redshifts \cite{lan2019luminosity}. In this case, the strength of the $z_{i}$ vs $z_{g}$ correlation decreases to $r = 0.33$, and the curves of about 20\% of LGRBs would not intersect the Yonetoku line, and hence have no solution. In the \(z_{\rm i}\) vs. \(z_{\rm g}\) plots shown in Figure \ref{fig:yonetoku2}, not only is the scatter around the \(z_{\rm } = z_{\rm g}\) line quite large, but the confidence intervals for individual  \(z_{i}\) are also very wide—exceeding 2.5 for GRBs with \(z_{\rm g} > 1.5\). For LGRBs, the 1\(\sigma\) confidence interval of $z_i$ (upper limit capped at 10) averages to 4.19, which is broader than the characteristic width (\(\Delta z \sim 3\)) of the metallicity-convolved GRB rate distribution. It is critical to emphasize again that this result is without considering for errors in measurements of fluence and peak energy. Such high uncertainties of the inferred $z_{\rm i}$ (significantly larger than the width of $z$ distribution of the population) calls into question the feasibility of any population studies which aim to infer distributions directly from pseudo-redshifts; for instance, it would be difficult to obtain any meaningful constraints on our understanding between the correlation of LGRBs and the cosmic star formation history, the characteristic delay time, or the luminosity function. \\}

\noindent In this work, we have assumed a k-correction factor of 1; in reality, the curves of  \( \mathcal{A}(z; E_{\mathrm{p,o}}, F) \) and \(\mathcal{Y}( z; E_{\mathrm{p,o}} , f_{\gamma} )\) would be slightly modified. \cite{zegarelli2022detection} found the k-correction distribution of GRBs detected by \textit{Fermi}-GBM to be highly skewed towards 1, with a median value of 1.12 and a mean value of 1.27, and with a maximum value of an outlier at \(\sim \) 4. Therefore, although in principle \(k(z)\) would increase with \(z\), the effect would be negligible due to the wide energy coverage of \textit{Fermi}-GBM. In fact, \cite{paul2018modelling} demonstrated that by assuming \textit{Fermi}-GBM’s GRBs follow the Band function and averaging the low- and high-energy indices \(\alpha\) and \(\beta\), the average spectral shape detected by  \textit{Fermi}-GBM yields a k-correction below 1.5 even at \(z = 10\). \\ 

{\emre 
\subsection{SGRBs}

\noindent   
Although most cosmological uses of GRBs focus on long-duration bursts, we have also briefly examined the case of SGRBs, whose redshift distribution is expected to peak at lower \(z\). With regards to the Amati relation, since the slope for  SGRBs $b_{\rm A} = 0.49 \pm 0.04$ is also below the limiting value for a unique solution, our conclusion for long GRBs extends equally to short GRBs in this context. With regards to the  Yonetoku relation, we sample the SGRBs from a BNS merger rate density (see appendix).  Assuming $\tau = 3$ Gyr, we obtain a correlation coefficient of $r = 0.48 \pm 0.04$ for \(z_{\rm i}\) vs. \(z_{\rm g}\), and the average 1 sigma confidence intervals of $z_i$ (capping the upper limit to 10 ) is 3.16, which greatly exceeds the delay time-scale ($\Delta z \sim 1$). In the case of $\tau$ = 1 Gyr, we find $r = 0.45 \pm 0.04$ for \(z_{\rm i}\) vs. \(z_{\rm g}\) and the average 1 sigma confidence intervals of $z_i$ is 3.57.  }

\begin{figure}[h]
    \centering
    \includegraphics[width=\linewidth]{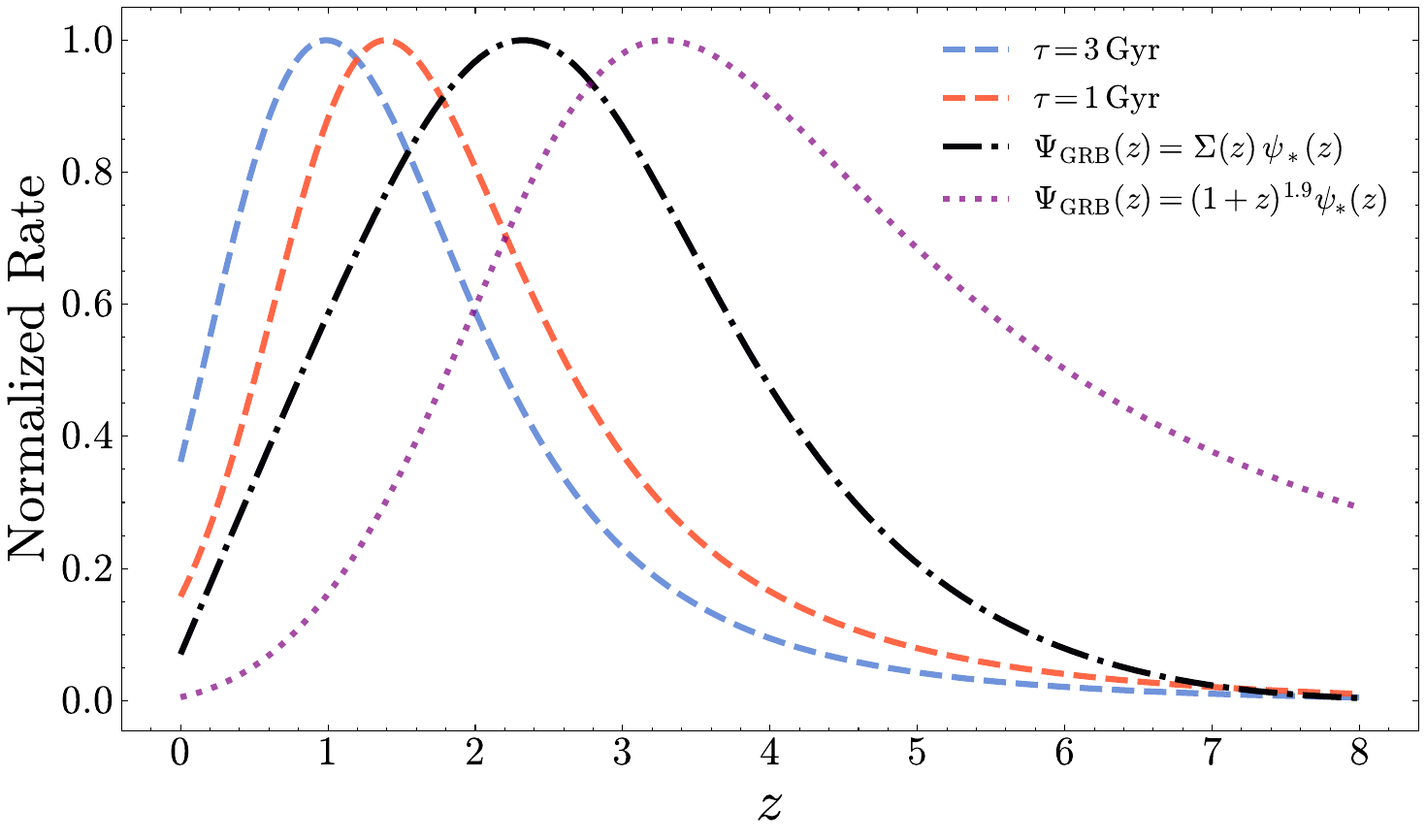}
    \caption{Normalized GRB rates utilized as a function of redshift. Blue: BNS merger rate with $\tau = 3$ Gyr; Red: BNS merger rate with $\tau = 1$ Gyr; Black: Metallicity-Convolved SFR rate; Purple: Density evolved SFR rate  }
    \label{fig:distributions}
\end{figure}

{\emre 
 \subsection{Concluding Remarks}

\noindent Our findings underscore the inherent difficulty in analytically solving for redshifts using phenomenological GRB correlations as distance indicators for both individual GRBs and population studies. Nevertheless, it is important to stress that these correlations, when combined with other observational information or advanced statistical/machine-learning frameworks, may still hold promise for constraining population distributions (e.g., luminosity functions, formation rates). For instance, Bayesian hierarchically informed methods or multi-parameter regressions leveraging detailed spectral shapes, afterglow properties, or prompt light-curve features could help refine pseudo-redshift estimates. In fact, recent studies have incorporated machine learning with great success in deriving pseudo-redshifts; \citep[See, for example,][]{dainotti2024inferring, aldowma2024deep}. In the study of \cite{aldowma2024deep}, for instance, the models included not only the key parameters of fluence, flux, and peak energy, but also additional spectral features, such as the low and high spectral indices of the Band function. This broader set of input variables allows machine learning approaches to identify complex, non-linear relationships that may not be immediately apparent in traditional regression-based methods. We stress that machine learning does not override the foundational limitations we present here, but instead provides a complementary way to address them by utilizing a wider range of observational data. Hence, we believe that the extensive data collected by \textit{Fermi}-GBM still retains potential in this regard. }

\noindent  We have not considered other phenomenological correlations of the prompt emission, such as \(E_{\mathrm{p},z} - E_{\mathrm{\gamma}}  \) (Ghirlanda) relation \citep{ghirlanda2004collimation},  \(E_{\mathrm{p},z} - L_{\mathrm{\gamma,p, iso }} - T_{0.45}   \) (Firmani) relation \citep{firmani2006discovery}, or those of the afterglow, such as the \(L_{\mathrm{X}} - T_{\mathrm{a}, z} \) (Dainotti) relation \citep{dainotti2008time} and \(E_{\mathrm{p},z} - E_{\mathrm{\gamma,iso }} - t_{\mathrm{b}, z}   \) (Liang-Zhang) relation \citep{liang2005model}. Nonetheless, if any such correlations are to be used as ``distance indicators", they must (i) yield a unique solution within a physical redshift range, and (ii) be tight enough to have a predictive power of statistical significance and yield reasonable error bands.

\begin{acknowledgments}
%\section{Acknowledgments}
ESY acknowledges support from the ``Alliance of International Science Organization (ANSO) Scholarship For Young Talents”. SXY acknowledges support from the Chinese Academy of Sciences (grant Nos. E32983U810 and E25155U110). SNZ acknowledges support from the National Natural Science Foundation of China (Grant No. 12333007 and 12027803) and International Partnership Program of Chinese Academy of Sciences (Grant No.113111KYSB20190020).

\end{acknowledgments}

\bibliography{ref}{}
\bibliographystyle{aasjournal}

\vspace{5mm}

\newpage 

\appendix

\section{Derivation of Tangent Redshift}

\noindent The Amati line is given by 
\begin{equation}
 \begin{aligned}
    \log E_{\mathrm{p}} = a_{\mathrm A} \log E_{\mathrm{iso}} + b_{\rm A}\,,
    \end{aligned}
\end{equation}
where \(a_{\mathrm A}\) is the slope of the Amati relation and \( b_{\rm A}\) is the intercept.  \( \mathcal{A}(z; E_{\mathrm{p,o}}, F) \) in \(E_{\mathrm{p}, z} - E_{\mathrm{iso}}\) space is parameterized by 
\begin{equation}
 \begin{aligned}
    E_{\mathrm{p}, z} = E_{\mathrm{p, o}} \times (1 + z)\,
 \end{aligned}
\end{equation}
\noindent and

\begin{equation}
 \begin{aligned}
    E_{\mathrm{iso}} = \frac{\mathrm{Fluence} \times 4\pi D^{2}_{\rm L}(z)}{1 + z}\,.
 \end{aligned}
\end{equation}

\noindent \textbf{Lemma}: The point on \( \mathcal{A}(z; E_{\mathrm{p,o}}, F) \) closest to the amati line {\emre for curves with no intersection}  occurs when the slope of \( \mathcal{A}(z; E_{\mathrm{p,o}}, F) \) in \(\log E_{{\rm p}, z} - \log E_{\mathrm{iso}}\) space matches the slope of the Amati line, \(a_{\mathrm A}\).\\

\noindent We first need to compute the derivatives of \( \log E_{\mathrm{p}}(z) \ \mathrm{and} \ \log E_{\mathrm{iso}}(z) \) with respect to \( \log(1 + z) \):

\begin{equation}
 \begin{aligned}
    \frac{d \log E_{\mathrm{p}}}{d \log(1+z)} =
     \frac{d \log E_{\mathrm{p, o}}}{d \log(1+z)} + 1
    \end{aligned}
\end{equation}\\
\noindent and 
\begin{equation}
 \begin{aligned}
    \frac{d \log E_{\mathrm{iso}}}{d \log(1+z)} = 
     2\frac{d \log D_{\rm L}(z)}{d \log(1+z)} - 1\,.
    \end{aligned}
\end{equation}

\noindent We need to find the redshift at which the Amati slope \(a_{\mathrm A}\) and the derivative of \( \mathcal{A}(z; E_{\mathrm{p,o}}, F) \) match: 
\begin{equation}
 \begin{aligned}
   \frac{d \log E_{\mathrm{p}, z}}{d \log E_{\mathrm{iso}}} = a_{\mathrm A}\,.
    \end{aligned}
\end{equation}\\

\begin{figure*}[h!]
    \centering
    \includegraphics[width=1\linewidth]{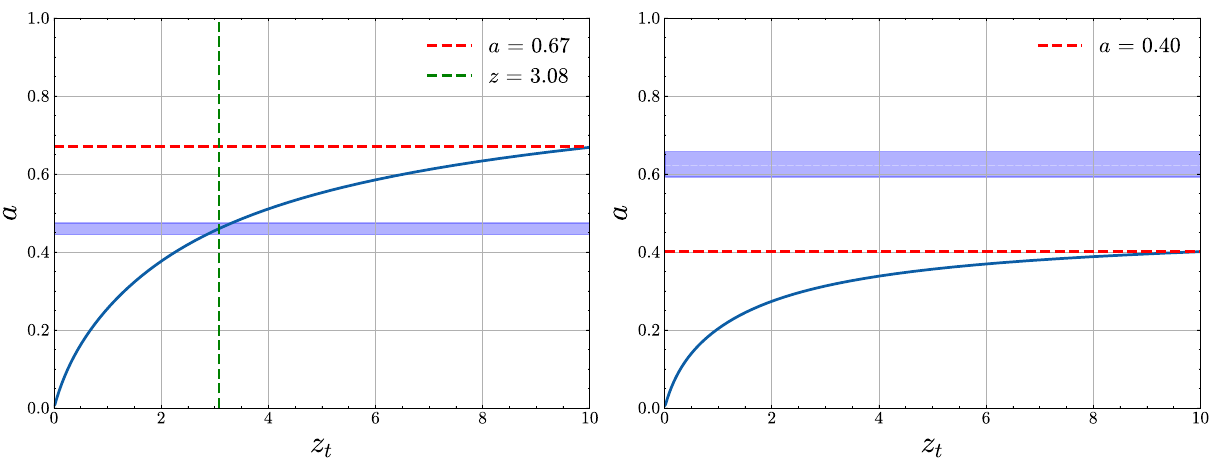}
    \caption{Plot of Equation \ref{eq: curve1} (left) and Equation \ref{eq: curve2} (right). The blue shaded region denotes the \(1\sigma\) uncertainty range for the slopes. The red dashed line denotes the limit above which the the function is well behaved (results in unique solution) for \( z \in [0.1, 10] \). The green dashed line corresponds to \(z_{\mathrm t}\) for the mean Amati slope. }
    \label{fig:limitingslope}
\end{figure*}

\noindent Hence:
\begin{equation}
 \begin{aligned}
 a_{\mathrm{A}} = \frac{\frac{d \log E_{\mathrm{p, o}}}{d \log(1+z)} + 1}{ 2 \frac{d \log D_{\mathrm{L}}(z)}{d \log(1+z)} - 1} = \frac{1}{2\frac{d \log D_{\mathrm{L}}(z_{\mathrm{t}})}{d \log(1+z_{\mathrm{t}})} - 1}\,
\end{aligned}
    \label{eq: curve1}
\end{equation}\\
since \(E_{\mathrm{p, o}}\) is a constant. It is easy to see that in the case of the Yonetoku relation, this becomes: 
\begin{equation}
 \begin{aligned}
   a_{\mathrm Y} = \frac{1}{2\frac{d \log D_{\mathrm{L}}(z_{\mathrm t})}{d \log(1+z_{\mathrm t})}}\,.
    \end{aligned}
    \label{eq: curve2}
\end{equation}\\
\newpage 
\noindent We plot Equation \ref{eq: curve1} in Figure \ref{fig:limitingslope}  (left), and  Equation \ref{eq: curve2} in Figure \ref{fig:limitingslope} (right). It is clear that \(z_{\mathrm t} = 3.08\) is at the midpoint of the intersection range between Equation \ref{eq: curve1} and the LGRB Amati slope uncertainty.  The Yonetoku and Amati relations results unique solutions within \(z \in [0.1, 10]\) for slopes which satisfy \(z_{\mathrm t} \geq 10 \). Clearly the observed slopes of both relations indicate that the Yonetoku relation is well-behaved (resulting unique solution), but the Amati relation is not.

%\section{Figures}

\begin{figure*}[h]
  \centering
  \begin{subfigure}[t]{0.5\textwidth} % Use nearly half the text width for each subfigure
    \centering
    \includegraphics[width=1.02\textwidth]{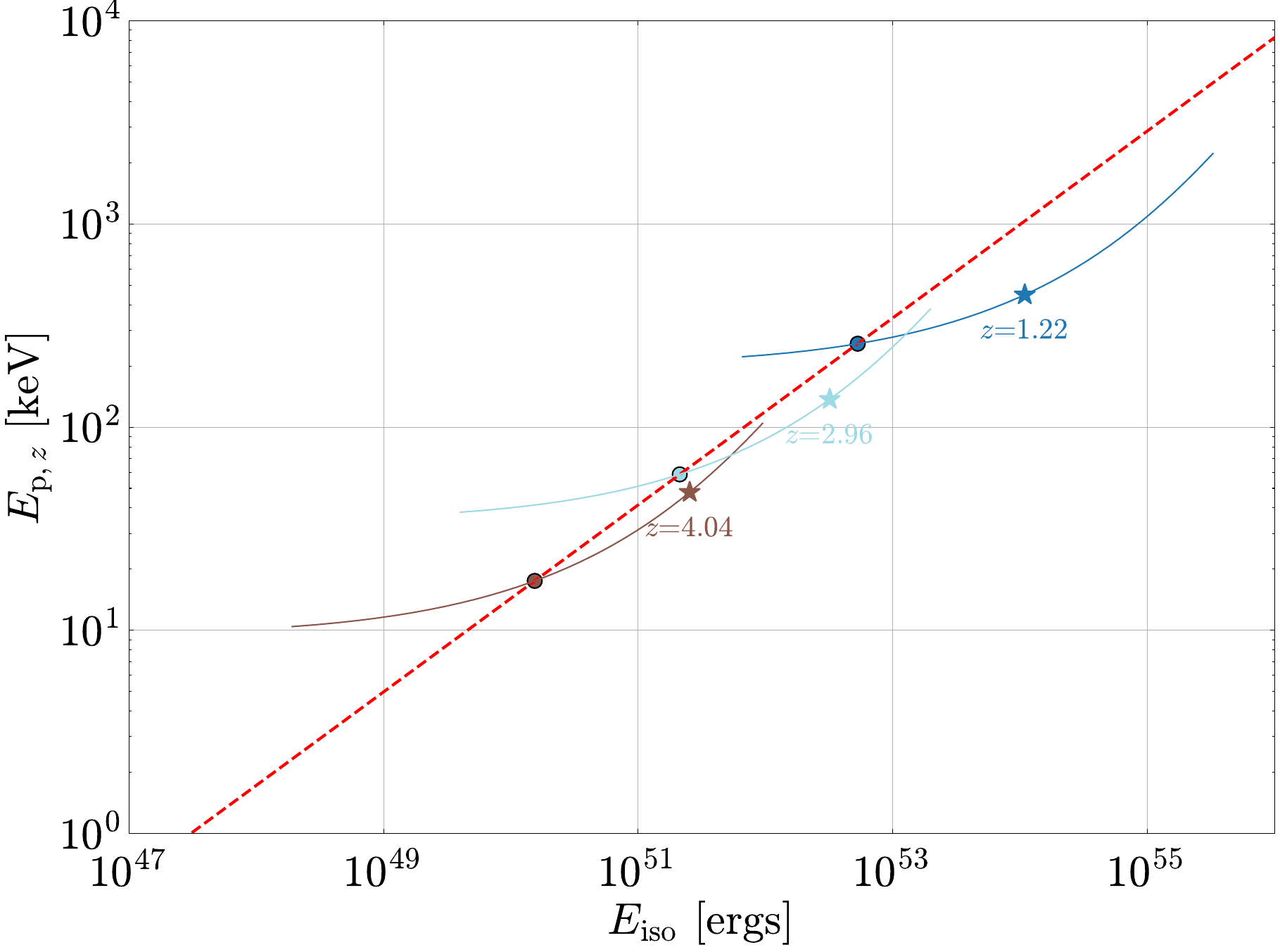} 
  \end{subfigure}
  %\hspace{0.001\textwidth}
  \begin{subfigure}[t]{0.48\textwidth} % Adjust width to fit perfectly next to each other
    \centering
    \includegraphics[width=0.94\textwidth]{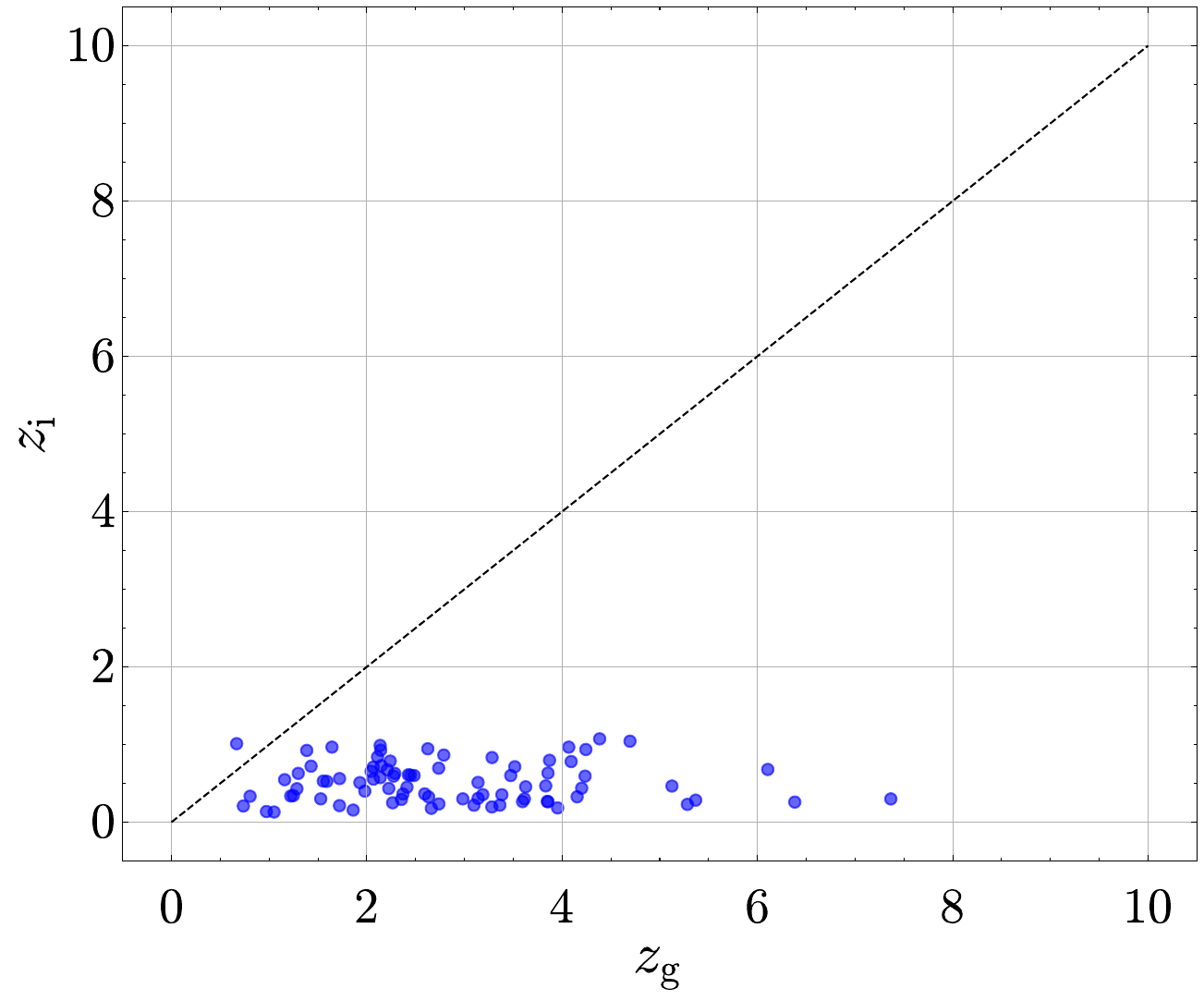}
  \end{subfigure}

  \caption{Plot of 3 sample \( \mathcal{A}(z; E_{\mathrm{p,o}}, F)\) curves with one solution only. Stars denote the true position of the GRB, with true redshifts annotated. \textbf{Right:} \(z_{\mathrm i}\) vs \(z_{\mathrm g}\) of 86 GRBs with one solution only (out of a sample of 250 GRBs)   }
  \label{fig:onesolution}
\end{figure*}

\noindent Figure \ref{fig:onesolution} illustrates the case where the curves  \( \mathcal{A}(z; E_{\mathrm{p,o}}, F) \) have only one solution when intrinsic scatter is introduced. This can be easily explained by noting that \(z_{\mathrm t}\) represents the highest possible redshift solution for curves that intersect the Amati line only once.

\newpage

\section{The redshift distribution when simulating the population of sGRBs}
\begin{figure*}[h]
  \centering
  \begin{subfigure}[t]{0.49\textwidth} % Use nearly half the text width for each subfigure
    \centering
    \includegraphics[width=0.97\textwidth]{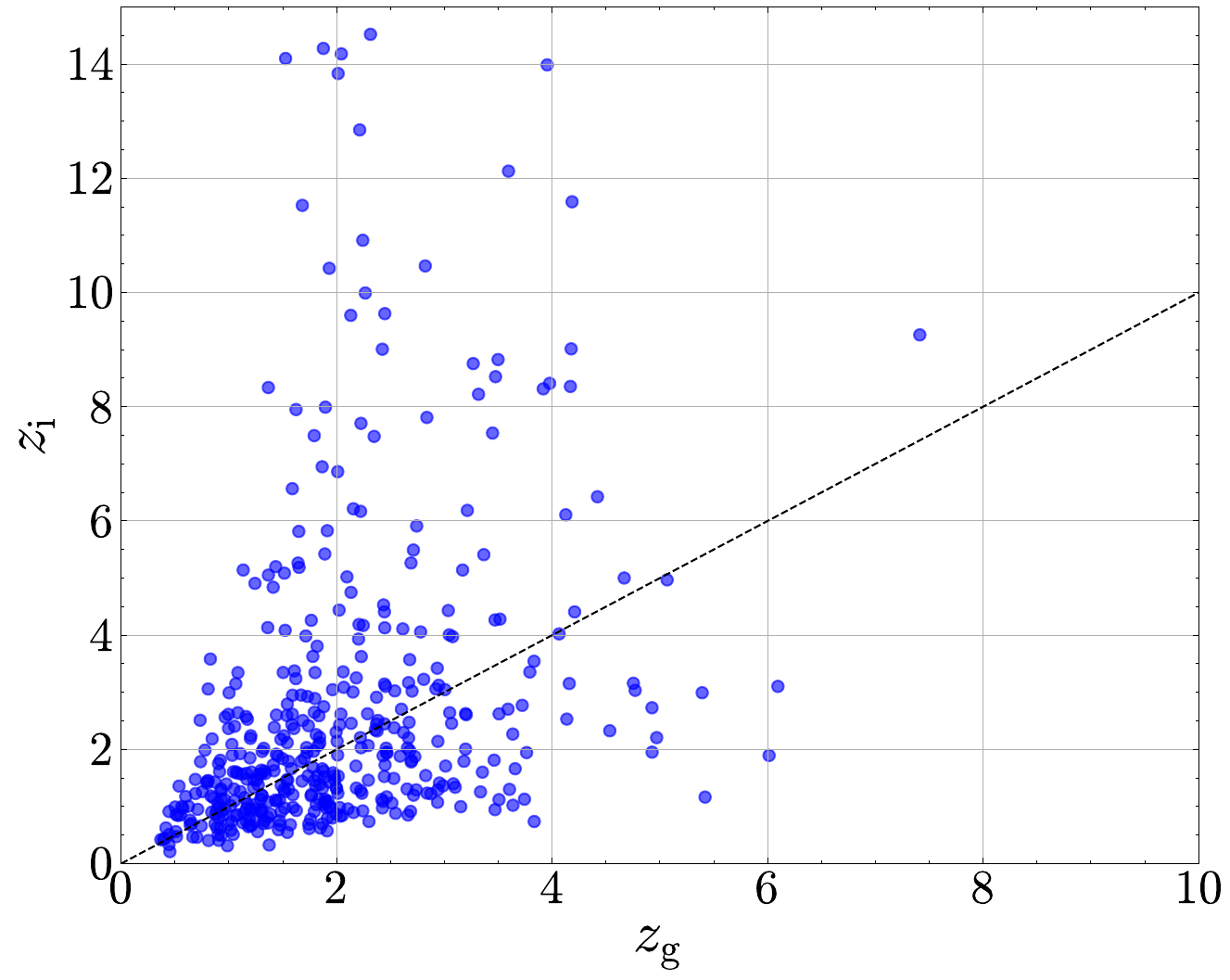} 
  \end{subfigure}
  %\hspace{0.001\textwidth}
  \begin{subfigure}[t]{0.48\textwidth} % Adjust width to fit perfectly next to each other
    \centering
    \includegraphics[width=1.02\textwidth]{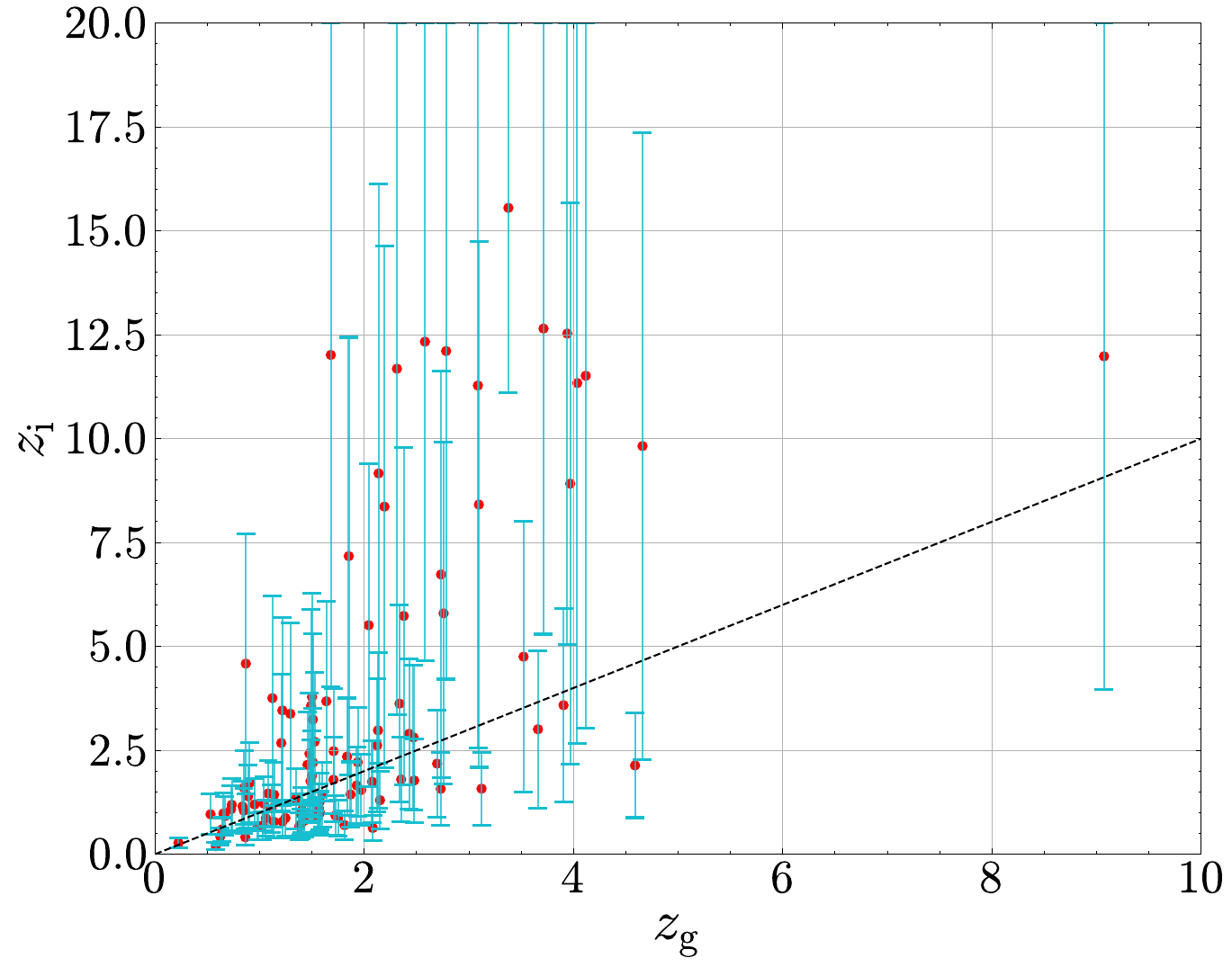}
  \end{subfigure}

  \caption{\textbf{Left:} \(z_{\mathrm i}\) vs \(z_{\mathrm g}\), with a Pearson Correlation Coefficient of $r = 0.48 \pm 0.04$ \href{https://code.ihep.ac.cn/emre/pseudo-redshifts/-/blob/main/Yonetoku\%20-\%20Scatter?ref\_type=heads}{\texttt{</>}} \textbf{Right: }Confidence intervals \([z_{\mathrm i}^{-\sigma}, z_{\mathrm i}^{+\sigma}]\) vs \(z_{\mathrm g}\) of  {\emre 100 } simulated SGRBs derived from the intersection of  \(\mathcal{Y}( z; E_{\mathrm{p,o}} , f_{\gamma} )\) with the Yonetoku uncertainty area \( \mathcal{U_{\mathrm Y}} \)}
  \label{fig:SGRB}
\end{figure*}

{\emre 

\noindent We assume the SGRBs to follow the Binary Neutron Star (BNS) merger rate. The differential BNS merger rate as function of redshift $R(z)\equiv \frac{dN}{dtdz}$, can be expressed in terms of the volumetric total BNS merger rate density $\mathcal{R}(z)\equiv \frac{dN}{dtdV_{c}}$ in the source frame as 
  $R(z)\equiv \frac{dN}{dtdz}$

\begin{equation}
R(z)=\frac{1}{1+z}\frac{dV_{c}}{dz}\mathcal{R}(z),
\label{r_r}
\end{equation}
where $dV_{c}/dz$ is the differential comoving volume. We adopt the Parametrization by \cite{vitale2019measuring} and take $\mathcal{R}(z)$ as:
\begin{equation}
\mathcal{R}(z_{\rm m})=\mathcal{R}_n\int_{z_{\rm m}}^{\infty}\psi(z_{\rm f})P(z_{\rm m}|z_{\rm f})dz_{\rm f},
\label{merger}
\end{equation}
where $\psi(z_{\rm f})$ is the non-normalized Madau-Dickinson star formation rate:
\begin{equation}
\psi(z)=\frac{(1+z)^{\alpha}}{1+(\frac{1+z}{C})^{\beta}},
\label{Madau}
\end{equation}
with $\alpha=2.7$, $\beta=5.6$, $C=2.9$ \citep{madau2014cosmic}, and $P(z_{\rm m}|z_{\rm f})$ is the probability that a BNS system merges at $z_{\rm m}$  given its formation at $z_{\rm f}$.  This is the distribution of delay times, which has the form: 
\begin{equation}
P(z_{\rm m}|z_{\rm f},\tau)=\frac{1}{\tau}{\rm exp}[-\frac{t_{\rm f}(z_{\rm f})-t_{\rm m}(z_{\rm m})}{\tau}]\frac{dt}{dz}.
\label{prob_merger}
\end{equation}

\noindent Here, $t_{\rm f}$ and $t_{\rm m}$ are the look back time as a function of $z_{\rm f}$ and $z_{\rm m}$, respectively. $\tau$ is the characteristic delay time. We sample SGRBs assuming a $\tau$ = 3 Gyr and 1 Gyr, and perform the same statistics as in section \ref{Yonetokuresults} . In Figure \ref{fig:SGRB}, we only show the results for the case of $\tau$ = 3 Gyr,  as it is skewed towards smaller redshifts compared to the case of $\tau$ = 1 Gyr (See Figure \ref{fig:distributions}).}

%\newpage



\end{document}